\documentclass[prd,superscriptaddress,twocolumn,nofootinbib,showpacs,preprintnumbers,floatfix]{revtex4}

\usepackage{mathrsfs}
\usepackage{amsfonts,amssymb,amsmath}
\usepackage{graphicx,epsfig}
\usepackage{multirow}

\begin{document}

\title{A Two-Tiered Correlation of Dark Matter with Missing Transverse Energy:
\\Reconstructing the Lightest Supersymmetric Particle Mass at the LHC}

\author{Tianjun Li}

\affiliation{State Key Laboratory of Theoretical Physics, Institute of Theoretical Physics,
Chinese Academy of Sciences, Beijing 100190, P. R. China }

\affiliation{George P. and Cynthia W. Mitchell Institute for Fundamental Physics and Astronomy,
Texas A$\&$M University, College Station, TX 77843, USA }

\author{James A. Maxin}

\affiliation{George P. and Cynthia W. Mitchell Institute for Fundamental Physics and Astronomy,
Texas A$\&$M University, College Station, TX 77843, USA }

\author{Dimitri V. Nanopoulos}

\affiliation{George P. and Cynthia W. Mitchell Institute for Fundamental Physics and Astronomy,
Texas A$\&$M University, College Station, TX 77843, USA }

\affiliation{Astroparticle Physics Group, Houston Advanced Research Center (HARC),
Mitchell Campus, Woodlands, TX 77381, USA}

\affiliation{Academy of Athens, Division of Natural Sciences,
28 Panepistimiou Avenue, Athens 10679, Greece }

\author{Joel W. Walker}

\affiliation{Department of Physics, Sam Houston State University,
Huntsville, TX 77341, USA }

%%%%%%%%%%%%%%%%%%%%%%%%%%%%%%%%%%%%%%%%%%%%%%%%%%%%%%%%%%%%%%%%%%%%%%%%%%%%

\begin{abstract}

We suggest that non-trivial correlations between the dark matter particle mass and collider based
probes of missing transverse energy $H_{\rm T}^{\rm miss}$ may facilitate a two tiered approach
to the initial discovery of supersymmetry and the subsequent reconstruction of the lightest supersymmetric
particle (LSP) mass at the LHC.  These correlations are demonstrated via extensive Monte Carlo simulation of
seventeen benchmark models, each sampled at five distinct LHC center-of-mass beam energies,
spanning the parameter space of No-Scale $\cal{F}$-$SU(5)$.  This construction is defined in turn by the union of the
${\cal F}$-lipped $SU(5)$ Grand Unified Theory, two pairs of hypothetical TeV scale vector-like supersymmetric
multiplets with origins in ${\cal F}$-theory, and the dynamically established boundary conditions of No-Scale Supergravity.
In addition, we consider a control sample comprised of a standard minimal Supergravity benchmark point.
Led by a striking similarity between the $H_{\rm T}^{\rm miss}$ distribution and the familiar power spectrum 
of a black body radiator at various temperatures, we implement a broad empirical fit of our simulation against
a Poisson distribution ans\"atz.  We advance the resulting fit as a theoretical blueprint for deducing the mass
of the LSP, utilizing only the missing transverse energy in a statistical sampling of $\ge$ 9 jet events.
Cumulative uncertainties central to the method subsist at a satisfactory $12$-$15\%$ level. 
The fact that supersymmetric particle spectrum of No-Scale $\cal{F}$-$SU(5)$ has thrived the withering onslaught of early LHC
data that is steadily decimating the Constrained Minimal Supersymmetric Standard Model and minimal Supergravity parameter
spaces is a prime motivation for augmenting more conventional LSP search methodologies with the presently proposed alternative.

\end{abstract}

\pacs{11.10.Kk, 11.25.Mj, 11.25.-w, 12.60.Jv}

\preprint{ACT-08-11, MIFPA-11-28}

\maketitle

%%%%%%%%%%%%%%%%%%%%%%%%%%%%%%%%%%%%%%%%%%%%%%%%%%%%%%%%%%%%%%%%%%%%%%%%%%%%

\section{Introduction}

%%%%%%%%%%%%%%%%%%%%%%%%%%%%%%%%%%%%%%%%%%%%%%%%%%%%%%%%%%%%%%%%%%%%%%%%%%%%

\subsection{Dark Matter and Missing Transverse Energy}

%%%%%%%%%%%%%%%%%%%%%%%%%%%%%%%%%%%%%%%%%%%%%%%%%%%%%%%%%%%%%%%%%%%%%%%%%%%%

High-precision experiments have revealed that the matter content of our Universe is
predominantly non-baryonic and dark.  The presence of large scale structure formation indicates that
dark matter should be primarily cold, {\it i.e.}~possessing a non-relativistic thermal speed at the time
of its own decoupling.  Measurements of the anisotropies in the Cosmic Microwave Background (CMB) radiation,
most recently and conclusively by the WMAP~\cite{Spergel:2003cb,Spergel:2006hy,Komatsu:2010fb} 
satellite experiment, allocate the proportional energy densities of baryonic matter, cold dark matter (CDM),
and dark energy (a cosmological constant) at about $4\%$, $23\%$, and $73\%$ respectively. 
A stable Weakly Interacting Massive Particle (WIMP) with a canonical weak-scale mass can naturally have a
thermal CDM relic density at the correct magnitude order~\cite{Ellis:1983ew}.
Although the Standard Model (SM) does not possess a CDM candidate, suitable particles
are manifest in various SM extensions, most notably in constructions featuring supersymmetry (SUSY),
which itself also efficiently addresses the quantum stability of the electroweak (EW) gauge hierarchy.
Decay of the Lightest Supersymmetric Particle (LSP), typically associated with a neutralino eigenstate,
may be blocked by a discrete $R$-parity symmetry, resulting in an exceedingly natural CDM candidate~\cite{Ellis:1983ew,Goldberg:1983nd}.
Alternative examples such as the little Higgs model with T-parity~\cite{Cheng:2003ju, Cheng:2004yc}
and extra dimensional models with KK parity~\cite{Servant:2002aq} are not considered further here.

In general, if dark matter particles are produced in a collider, being $U(1)_{\rm EM}$ neutral, $SU(3)_{\rm C}$ singlets,
they will escape the detector without direct observation.  The sole evidence of their production and rapid departure
will be that of a deficit in the zero sum of momentum over all observed particles with a component trajectory
which is perpendicular to the beamline.  This so-called ``missing transverse energy'' ($H_T^{\rm miss}$) is measurable to
fine precision at the LHC, which has been steadily accumulating data, thus far without any positively reported evidence for supersymmetry, from
${\sqrt s}=7$ TeV proton-proton collisions since March 2010~\cite{Khachatryan:2011tk, daCosta:2011hh, daCosta:2011qk}.
In the current early operational stage of the LHC, our attention has unquestionably been piqued by the relentless severe reductions
inflicted upon the viable minimal Supergravity (mSUGRA) and Constrained Minimal Supersymmetric Standard Model (CMSSM)
parameter space.  It was concluded by one particular study that the miniscule $35~{\rm pb}^{-1}$ of initially reported
luminosity was alone sufficient to impeach more than $99\%$ of the parameter space~\cite{Strumia:2011}, although other analysts
remained more circumspect~\cite{Buchmueller:2011ki}.  The negative results have now held through the collection of more
than a femtobarn of integrated luminosity by each of the ATLAS~\cite{Aad:2011qa} and CMS~\cite{PAS-SUS-11-003} collaborations,
although we must remark that tantalizing low-statistics excesses do exist in both cases, which are neatly accounted for
by simulation of the No-Scale $\cal{F}$-$SU(5)$ collider-detector signal~\cite{Li:2011av}.  Regardless, should this
situation much longer persist into the deconstruction of the burgeoning wealth of collected raw data,
reported to have already reached the impressive milestone of $5~{\rm fb}^{-1}$ by the close of 2011,
then the first great question of the LHC era shall become significantly heightened.  Namely, that is the question of
whether there exist specifically detailed SUSY and/or superstring based post-SM constructions which
simultaneously escape the existing experimental onslaught, yet also remain in principle discoverable
to the present generation collider.

Granting, momentarily, an answer in the tacit affirmative to the prior question of discovery, one may speculate
regarding the equally great question immediately subsequent, that being the issue of the detailed SUSY CDM classification.
Differentiation between contending scenarios, both within the CMSSM and without, will require a probe at present and future
collider and direct detection experiments, the latter now led by the XENON100 collaboration~\cite{Aprile:2011hi},
of the intrinsic properties of the CDM, such as its spin and mass, and the structure of its interaction with SM particles.
Assuming a SUSY LSP to represent the CDM, we may in principle make a collider based measurement of the LSP mass from the decay
chains of heavier particles, for example linking the squark and gluino to the lightest neutralino in the supersymmetric
SM~\cite{Polesello:2004qy, Nojiri:2005ph}, though this method is highly model dependent, and there are myriad complications
with its interpretation.  Firstly, since the momentum component along beamline can by its nature never be collected,
it is impossible to reconstruct the true net missing energy total.  Secondly, $R$-parity implies that the LSP candidate
will always ultimately appear pairwise.  Moreover, the two neutralinos may be kinematically decoupled from the original
back-to-back pair production event, {\it e.g.} of two heavy squarks $\widetilde{q}\,\widetilde{\overline{q}}$, by an
intricate decay cascade, such that the final relative angular orientation is purely random, allowing for any possible partial cancellation.
Finally, the dark matter component of the total missing energy $H_T^{\rm miss}$ must be segregated from other
potential sources, such as energetic neutrinos and energy mismeasurement, for example by particle escape through a
calorimeter seam.  Although the prospect of a substantial loss of the latter type is unlikely, it competes with the ever
more extraordinarily rare target events.  These foreboding circumstances compel some candid apprehension against
the hope that any fundamental correlation might endure between the dark matter particle mass and the missing transverse
energy measurement.

In the present work we set out to directly confront, by the aid of a massive Monte Carlo collider and detector simulation, the
two great questions here posed.  The two tiered nature of our presentation, approaching first the prospect of SUSY CDM
discovery, and secondly the subsequent classification of the SUSY CDM signal via explicit numerical correlation of the LSP mass
with the peak of the missing transverse energy distribution, is in correspondence with our view of how the
experimental reality might plausibly unfold in the coming months and years.
Our study takes the context of a specific model, dubbed No-Scale $\cal{F}$-$SU(5)$~\cite{
Li:2010ws, Li:2010mi,Li:2010uu,Li:2011dw, Li:2011hr, Maxin:2011hy,Li:2011xu, Li:2011in,Li:2011rp,Li:2011fu,Li:2011ex,Li:2011av,Li:2011ab},
representing the merger of the ${\cal F}$-lipped $SU(5)$ Grand Unified Theory
(GUT)~\cite{Barr:1981qv,Derendinger:1983aj,Antoniadis:1987dx},
two pairs of hypothetical TeV scale vector-like supersymmetric multiplets with origins in
${\cal F}$-theory~\cite{Jiang:2006hf,Jiang:2009zza,Jiang:2009za,Li:2010dp,Li:2010rz},
and the dynamically established boundary conditions of No-Scale
Supergravity~\cite{Cremmer:1983bf,Ellis:1983sf, Ellis:1983ei, Ellis:1984bm, Lahanas:1986uc}.
The construction features an essential phenomenological consistency,
profound predictive capacity, a singularly distinctive experimental signature, and
imminent testability.  The core of the model space remains experimentally unblemished, in stark contrast to
the incisive cuts already made into the heart of the more traditional mSUGRA/CMSSM formulations.
This deft evasion of all existing LEP, Tevatron, and LHC sparticle mass limits stems from the characteristic sparticle mass
hierarchy $m_{\tilde{t}_1} < m_{\tilde{g}} < m_{\tilde{q}}$ of a light stop and gluino, both comfortably lighter than all other
squarks, a feature which is stable across the full model space.
If the tide advancing against the CMSSM should continue unabated, then island refuges such as
No-Scale $\cal{F}$-$SU(5)$ will demand increasing deference in the search for
a physically feasible supersymmetric grand unified theory.

%%%%%%%%%%%%%%%%%%%%%%%%%%%%%%%%%%%%%%%%%%%%%%%%%%%%%%%%%%%%%%%%%%%%%%%%%%%%

\subsection{A No-Scale ${\cal F}$-$SU(5)$ Primer}

%%%%%%%%%%%%%%%%%%%%%%%%%%%%%%%%%%%%%%%%%%%%%%%%%%%%%%%%%%%%%%%%%%%%%%%%%%%%

The No-Scale $\cal{F}$-$SU(5)$ construction inherits all of the most beneficial phenomenology~\cite{Nanopoulos:2002qk}
of flipped $SU(5)$~\cite{Barr:1981qv,Derendinger:1983aj,Antoniadis:1987dx},
including fundamental GUT scale Higgs representations (not adjoints), natural doublet-triplet
splitting, suppression of dimension-five proton decay and a two-step see-saw mechanism
for neutrino masses, as well as all of the most beneficial theoretical motivation of No-Scale
Supergravity~\cite{Cremmer:1983bf,Ellis:1983sf, Ellis:1983ei, Ellis:1984bm, Lahanas:1986uc},
including a deep connection to the string theory infrared limit
(via compactification of the weakly coupled heterotic theory~\cite{Witten:1985xb} or 
M-theory on $S^1/Z_2$ at the leading order~\cite{Li:1997sk}),
the natural incorporation of general coordinate invariance (general relativity),
a mechanism for SUSY breaking which preserves a vanishing cosmological constant at the tree level
(facilitating the observed longevity and cosmological flatness of our Universe~\cite{Cremmer:1983bf}),
natural suppression of CP violation and flavor-changing neutral currents, dynamic stabilization
of the compactified spacetime by minimization of the loop-corrected scalar potential and a dramatic
reduction in parameterization freedom.

Written in full, the gauge group of flipped $SU(5)$ is $SU(5)\times U(1)_{X}$, which can be embedded into $SO(10)$.
The generator $U(1)_{Y'}$ is defined for fundamental five-plets as $-1/3$ for the triplet members, and $+1/2$ for the doublet.
The hypercharge is given by $Q_{Y}=( Q_{X}-Q_{Y'})/5$.  There are three families of Standard Model (SM) fermions,
whose quantum numbers under the $SU(5)\times U(1)_{X}$ gauge group are
\begin{equation}
F_i={\mathbf{(10, 1)}} \quad;\quad {\bar f}_i={\mathbf{(\bar 5, -3)}} \quad;\quad {\bar l}_i={\mathbf{(1, 5)}},
\label{eq:smfermions}
\end{equation}
where $i=1, 2, 3$.  There is a pair of ten-plet Higgs for breaking the GUT symmetry, and a pair
of five-plet Higgs for electroweak symmetry breaking (EWSB).
\begin{eqnarray}
& H={\mathbf{(10, 1)}}\quad;\quad~{\overline{H}}={\mathbf{({\overline{10}}, -1)}} & \nonumber \\
& h={\mathbf{(5, -2)}}\quad;\quad~{\overline h}={\mathbf{({\bar {5}}, 2)}} &
\label{eq:Higgs}
\end{eqnarray}
Since we do not observe mass degenerate superpartners for the known SM fields, SUSY must itself be broken around the TeV scale.
In the minimal supergravities (mSUGRA), this occurs first in a hidden sector, and the secondary propagation by gravitational interactions
into the observable sector is parameterized by universal SUSY-breaking ``soft terms'' which include the gaugino mass $M_{1/2}$, scalar mass
$M_0$ and the trilinear coupling $A$.  The ratio of the low energy Higgs vacuum expectation values (VEVs) $\tan \beta$, and the sign of
the SUSY-preserving Higgs bilinear mass term $\mu$ are also undetermined, while the magnitude of the $\mu$ term and its bilinear soft term $B_{\mu}$
are determined by the $Z$-boson mass $M_Z$ and $\tan \beta$ after EWSB.  In the simplest No-Scale scenario,
$M_0$=A=$B_{\mu}$=0 at the unification boundary, while the complete collection of low energy SUSY breaking soft-terms evolve down 
with a single non-zero parameter $M_{1/2}$.  Consequently, the particle spectrum will be proportional to $M_{1/2}$ at leading order,
rendering the bulk ``internal'' physical properties invariant under an overall rescaling.
The rescaling symmetry can likewise be broken to a certain degree by the vector-like
mass parameter $M_{\rm V}$, although this effect is weak.

The matching condition between the low-energy value of
$B_\mu$ that is demanded by EWSB and the high-energy $B_\mu = 0$ boundary is notoriously difficult to reconcile under the
renormalization group equation (RGE) running.  The present solution relies on modifications to the $\beta$-function coefficients that are generated by the inclusion
of the extra vector-like multiplets, which may actively participate in radiative loops above their characteristic mass threshold $M_{\rm V}$. 
Naturalness in view of the gauge hierarchy and $\mu$ problems suggests that the mass $M_{\rm V}$ should be of the TeV order.
Avoiding a Landau pole for the strong coupling constant restricts the set of vector-like multiplets which may be
given a mass in this range to only two constructions with flipped charge assignments, which have been explicitly realized
in the $F$-theory model building context~\cite{Jiang:2006hf,Jiang:2009zza, Jiang:2009za}.  We adopt the multiplets
\begin{eqnarray}
& {XF} \equiv {\mathbf{(10,1)}} = (XQ,XD^c,XN^c) ~;~ {\overline{XF}} \equiv {\mathbf{({\overline{10}},-1)}} & \nonumber \\
& {Xl} = {\mathbf{(1, -5)}} \quad;\quad{\overline{Xl}} = {\mathbf{(1, 5)}}\equiv XE^c \, , &
\label{eq:flippons}
\end{eqnarray}
where $XQ$, $XD^c$, $XE^c$ and $XN^c$ carry the same quantum numbers as the quark doublet, right-handed down-type quark,
charged lepton and neutrino, respectively.  Alternatively, the pair of $SU(5)$ singlets may be discarded, but phenomenological consistency then
requires the substantial application of unspecified GUT thresholds.  In either case, the (formerly negative) one-loop $\beta$-function
coefficient of the strong coupling $\alpha_3$ becomes precisely zero, flattening the RGE running, and generating a wide
gap between the large $\alpha_{32} \simeq \alpha_3(M_{\rm Z}) \simeq 0.11$ and the much smaller $\alpha_{\rm X}$ at the scale $M_{32}$ of the intermediate
flipped $SU(5)$ unification of the $SU(3) \times SU(2)_{\rm L}$ subgroup.  This facilitates a very significant secondary running phase
up to the final $SU(5) \times U(1)_{\rm X}$ unification scale~\cite{Li:2010dp}, which may be elevated by 2-3 orders of magnitude
into adjacency with the Planck mass, where the $B_\mu = 0$ boundary condition fits like hand to glove~\cite{Ellis:2001kg,Ellis:2010jb,Li:2010ws}.  
This natural resolution of the ``little hierarchy'' problem corresponds also to true string-scale gauge coupling unification in
the free fermionic string models~\cite{Jiang:2006hf,Lopez:1992kg} or the decoupling scenario in F-theory models~\cite{Jiang:2009zza,Jiang:2009za},
and also helps to address the monopole problem via hybrid inflation.

The modifications to the $\beta$-function coefficients from introduction of the vector-like multiplets have a
parallel effect on the RGEs of the gauginos.  In particular, the color-charged gaugino mass $M_{\rm 3}$ likewise runs down flat from the
high energy boundary, obeying the relation $M_3/M_{1/2} \simeq \alpha_3(M_{\rm Z})/\alpha_3(M_{32}) \simeq \mathcal{O}\,(1)$,
which precipitates a conspicuously light gluino mass assignment.
The $SU(2)_{\rm L}$ and hypercharge $U(1)_{\rm Y}$ associated gaugino masses are by
contrast driven downward from the $M_{1/2}$ boundary value by roughly the ratio of their corresponding gauge couplings
$(\alpha_2,\alpha_{\rm Y})$ to the strong coupling $\alpha_{\rm s}$.
The large mass splitting expected from the heaviness of the top quark via its strong coupling to the Higgs (which is also key to generating an appreciable
radiative Higgs mass shift $\Delta~m_h^2$~\cite{Li:2011ab}) is responsible for a rather light stop squark $\widetilde{t}_1$.
The distinctively predictive $m_{\tilde{t}_1} < m_{\tilde{g}} < m_{\tilde{q}}$ mass hierarchy of a light stop
and gluino, both much lighter than all other squarks, is stable across the full No-Scale $\cal{F}$-$SU(5)$ model space, but is
not precisely replicated in any phenomenologically favored constrained MSSM (CMSSM) constructions of which we are aware.

This spectrum generates a unique event topology starting from the pair production of heavy squarks
$\widetilde{q} \widetilde{\overline{q}}$, except for the light stop, in the initial hard scattering process,
with each squark likely to yield a quark-gluino pair $\widetilde{q} \rightarrow q \widetilde{g}$.  Each gluino may be expected
to produce events with a high multiplicity of virtual stops, via the (possibly off-shell) $\widetilde{g} \rightarrow \widetilde{t}$
transition, which in turn may terminate into hard scattering products such as $\rightarrow W^{+}W^{-} b \overline{b} \widetilde{\chi}_1^{0}$
and $W^{-} b \overline{b} \tau^{+} \nu_{\tau} \widetilde{\chi}_1^{0}$, where the $W$ bosons will produce mostly hadronic jets and some leptons.
The model described may then consistently exhibit a net product of eight or more hard jets emergent from a single squark pair production event,
passing through a single intermediate gluino pair, resulting after fragmentation in a spectacular signal of ultra-high multiplicity final state jet events.
We remark also that the entirety of the viable $\cal{F}$-$SU(5)$ parameter space naturally features a
dominantly Bino LSP, at a purity greater than 99.7\%, as is exceedingly suitable for direct detection, for example by
XENON100~\cite{Aprile:2011hi,Li:2011in}.  There exists no direct Bino to Wino mass mixing term.
This distinctive and desirable model characteristic is guaranteed by the relative heaviness of the Higgs bilinear
mass $\mu$, which in the present construction generically traces the universal gaugino mass $M_{1/2}$ at the boundary
scale $M_{\cal F}$, and subsequently transmutes under the RGEs to a somewhat larger value at the electroweak scale.

A majority of the bare-minimally constrained~\cite{Li:2011xu} parameter space of No-Scale $\cal{F}$-$SU(5)$,
as defined by consistency with the world average top-quark mass $m_{\rm t}$, the No-Scale boundary conditions,
radiative EWSB, the centrally observed WMAP7 CDM relic density~\cite{Komatsu:2010fb}, and precision LEP constraints on the
lightest CP-even Higgs boson $m_{h}$~\cite{Barate:2003sz,Yao:2006px} and other light SUSY chargino and neutralino mass content,
remains viable even after careful comparison against the first inverse femtobarn of LHC data~\cite{Li:2011fu,Li:2011av}.
Moreover, a highly favorable ``golden'' subspace~\cite{Li:2010ws,Li:2010mi,Li:2011xg} exists which may simultaneously account for the key rare process
limits on the muon anomalous magnetic moment $(g~-~2)_\mu$ and the branching ratio of the flavor-changing neutral current decays
$b \to s\gamma$ and $B_{s}^{0} \to \mu^+\mu^-$.  The intersection of these experimental bounds is highly non-trivial,
as the tight theoretical constraints, most notably the vanishing of $B_\mu$ at the high scale boundary, render the residual
parameterization deeply insufficient for arbitrary tuning of even isolated predictions, let alone the union of all predictions.
In addition, a top-down consistency condition on the gaugino boundary mass $M_{1/2}$, and the parametrically coupled value of $\tan\beta$,
is dynamically determined at a secondary local minimization $dV_{\rm min}/dM_{1/2}=0$ of the minimum of the Higgs potential $V_{\rm min}$;
since $M_{1/2}$ is related to the modulus field of the internal string theoretic space, this also
represents a dynamic stabilization of the compactification modulus.  The result is demonstrably consistent with the bottom-up
phenomenological approach ~\cite{Li:2010uu,Li:2011dw,Li:2011xu,Li:2011ex}, and we note in particular a rather distinctive conclusion
that is enforced in both perspectives: the ratio $\tan\beta$ must have a value very close to 20.

%%%%%%%%%%%%%%%%%%%%%%%%%%%%%%%%%%%%%%%%%%%%%%%%%%%%%%%%%%%%%%%%%%%%%%%%%%%%

\section{Tier 1: SUSY CDM Discovery}

%%%%%%%%%%%%%%%%%%%%%%%%%%%%%%%%%%%%%%%%%%%%%%%%%%%%%%%%%%%%%%%%%%%%%%%%%%%%

\begin{table*}[htbp]
	\centering
	\caption{We tabulate various statistics for seventeen representative points selected from the viable No-Scale $\cal{F}$-$SU(5)$ parameter
	space, each satisfying the bare-minimal phenomenological constraints outlined in Ref.~\cite{Li:2011xu}.
	The dimensionful parameters $m_{\rm LSP}, M_{1/2}, M_{\rm V}, m_{t}, B_{\mu}(M_{\cal F}), \mu(M_{\rm Z})$, and $m_{\chi^{0}_{1}}$ are in units of GeV.
	The discoverability statistic $S/\sqrt{B+1}$ is computed at a $\sqrt{s}=7$~TeV LHC, scaled to $1~{\rm fb}^{-1}$ of
	integrated luminosity, employing the {\tt CMS HYBRID} selection cut methodology~\cite{Li:2011av} in our own Monte Carlo,
	and adopting a corresponding SM background of $B = 3.1$ events, as extracted from a report by the CMS collaboration~\cite{PAS-SUS-11-003}.}
		\begin{tabular}{c|c|c|c|c|c|c|c|c|c|c} \hline
$~~	m_{\rm LSP}	~~$&$~~	M_{1/2}	~~$&$	~~M_{\rm V}	~~$&$	~~\tan\beta~~	$&$	~~~m_{\rm t}~~~	$&$	~~B_{\mu}(M_{\cal{F}})~~
	$&$	~~\mu(M_{Z})~~	$&$	~~~~\Omega_{\chi^{0}_{1}}~~~~	$&$	\sigma_{\rm SI}~(\times10^{-10}{\rm pb})
	$&$~~~	m_{\chi^{0}_{1}}~~~	$&$~~~  S/\sqrt{B+1}~~~	$	\\	\hline	\hline
$	74.8	$&$	385	$&$	3575	$&$	19.8	$&$	172.5	$&$	-0.60	$&$	644	$&$	0.1117	$&$	6.80	$&$	74.8	$&$	5.8	$	\\	\hline	
$	75.0	$&$	395	$&$	2075	$&$	19.7	$&$	172.5	$&$	0.96	$&$	679	$&$	0.1106	$&$	5.20	$&$	75.0	$&$	6.2	$	\\	\hline	
$	74.7	$&$	400	$&$	1450	$&$	19.5	$&$	173.7	$&$	-0.96	$&$	706	$&$	0.1157	$&$	4.20	$&$	74.7	$&$	5.5	$	\\	\hline	
$	75.0	$&$	410	$&$	925	  $&$	19.4	$&$	174.4	$&$	0.92	$&$	744	$&$	0.1123	$&$	3.20	$&$	75.0	$&$	4.9	$	\\	\hline	
$	83.2	$&$	425	$&$	3550	$&$	20.4	$&$	172.2	$&$	-0.73	$&$	700	$&$	0.1117	$&$	4.70	$&$	83.2	$&$	4.7	$	\\	\hline	
$	83.1	$&$	435	$&$	2000	$&$	20.1	$&$	173.1	$&$	0.62	$&$	743	$&$	0.1131	$&$	3.40	$&$	83.1	$&$	4.0	$	\\	\hline	
$	82.8	$&$	445	$&$	1125	$&$	19.9	$&$	174.4	$&$	0.58	$&$	789	$&$	0.1101	$&$	2.50	$&$	82.8	$&$	3.5	$	\\	\hline	
$	92.2	$&$	465	$&$	3850	$&$	20.7	$&$	172.2	$&$	0.61	$&$	755	$&$	0.1146	$&$	3.30	$&$	92.2	$&$	3.7	$	\\	\hline	
$	92.2	$&$	475	$&$	2400	$&$	20.6	$&$	173.1	$&$	0.55	$&$	793	$&$	0.1146	$&$	2.60	$&$	92.2	$&$	3.3	$	\\	\hline	
$	92.1	$&$	485	$&$	1475	$&$	20.4	$&$	174.3	$&$	0.14	$&$	836	$&$	0.1092	$&$	1.90	$&$	92.1	$&$	3.0	$	\\	\hline	
$	100.5	$&$	505	$&$	3700	$&$	21.0	$&$	172.6	$&$	0.65	$&$	818	$&$	0.1109	$&$	2.30	$&$	100.5	$&$	2.8	$	\\	\hline	
$	100.2	$&$	510	$&$	2875	$&$	21.0	$&$	174.1	$&$	0.64	$&$	840	$&$	0.1122	$&$	2.00	$&$	100.2	$&$	2.4	$	\\	\hline	
$	100.0	$&$	520	$&$	1725	$&$	20.7	$&$	174.4	$&$	-0.19	$&$	882	$&$	0.1144	$&$	1.60	$&$	100.0	$&$	1.9	$	\\	\hline	
$	108.8	$&$	560	$&$	1875	$&$	21.0	$&$	174.4	$&$	0.94	$&$	937	$&$	0.1131	$&$	1.20	$&$	108.8	$&$	1.4	$	\\	\hline	
$	133.2	$&$	650	$&$	4700	$&$	22.0	$&$	173.4	$&$	0.19	$&$	1009	$&$	0.1115	$&$	0.96	$&$	133.2	$&$	0.6	$	\\	\hline	
$	155.9	$&$	750	$&$	5300	$&$	22.5	$&$	174.4	$&$	0.46	$&$	1143	$&$	0.1118	$&$	0.59	$&$	155.9	$&$	0.1	$	\\	\hline	
$	190.5	$&$	900	$&$	6000	$&$	23.0	$&$	174.4	$&$	0.63	$&$	1335	$&$	0.1120	$&$	0.33	$&$	190.5	$&$	0.0	$	\\	\hline	
		\end{tabular}
		\label{tab:points}
\end{table*}

%%%%%%%%%%%%%%%%%%%%%%%%%%%%%%%%%%%%%%%%%%%%%%%%%%%%%%%%%%%%%%%%%%%%%%%%%%%%

\subsection{Missing Transverse Energy as a SUSY Signal\label{sct:MET}}

%%%%%%%%%%%%%%%%%%%%%%%%%%%%%%%%%%%%%%%%%%%%%%%%%%%%%%%%%%%%%%%%%%%%%%%%%%%%

In the first tier of this study, we present an exhaustive Monte Carlo simulation of seventeen benchmark points, as detailed in
Table~\ref{tab:points}.  Three heavier benchmarks have been chosen here to supplement the original fourteen from Ref.~\cite{Li:2011xu}.
Together, these sample points span the parameter space of bare minimal phenomenological constraints~\cite{Li:2011xu} on No-Scale $\cal{F}$-$SU(5)$.
We also consider the third ``Snowmass Points and Slopes'' (SPS) benchmark point (SPS3)~\cite{Allanach:2002nj}
as a well established mSUGRA control sample, with $m_{\rm LSP}=161.7$~GeV, $M_0 = 90$~GeV, $M_{1/2}=400$~GeV, $m_{\rm t}=175$~GeV and $\tan \beta = 10$.
We present an example SUSY spectrum for the favorable $m_{\rm LSP}=100.0$~GeV, $M_{1/2}=520$~GeV benchmark in Table~\ref{tab:masses}.

A major new feature of the current study is the extension of our prior simulation
coverage to include five different LHC center-of-mass beam energies, starting from the presently operational $\sqrt{s} = 7$~TeV machine,
and culminating in the bright future of a $\sqrt{s} = 14$~TeV LHC.  Our initial purpose here is the development of a comprehensive map of the prospects
for SUSY discovery across this model and energy space, based on the raw count of events surviving a set of cuts designed for suppression of SM backgrounds.
The entirety of our collider based analysis is in fact dependent upon adoption of an ultra-high multiplicity
$(\ge 9)$ jet event cutting methodology~\cite{Li:2011hr, Maxin:2011hy}.  The heart of this approach is the observation that the decay chains
described in the introduction, which are prevalent for the signature $\cal{F}$-$SU(5)$ mass hierarchy, will very frequently result
in at least eight hard jets per SUSY event (the $W$ bosons will produce mostly hadronic jets and some leptons), and given fragmentation
processes, perhaps a much larger number of softer final state jets.  Since the ultra-high jet regime has been demonstrated~\cite{Maxin:2011hy} to
be naturally suppressed in the SM backgrounds, we are further able to relax certain of the harsh cuts which are very effective for separating
out the CMSSM in intermediate jet searches, but which simultaneously exert a costly attrition against our signal.

\begin{table}[htbp]
  \small
        \centering
        \caption{The detailed SUSY particle spectrum (in GeV) for the experimentally favorable benchmark
        with $M_{1/2}$ = 520 GeV, $\tan \beta = 20.7$, $M_{V}$ = 1725, and $m_{t}$ = 174.4 GeV.  Note that
	the presented light CP-even Higgs mass $m_h$ does not include radiative corrections from the vector-like
	``flippon'' multiplets, which may account for an upward shift of around $3$-$4$~GeV~\cite{Li:2011ab}.}
                \begin{tabular}{|c|c||c|c||c|c||c|c||c|c||c|c|} \hline
    $\widetilde{\chi}_{1}^{0}$&$100$&$\widetilde{\chi}_{1}^{\pm}$&$217$&$\widetilde{e}_{R}$&$196$&$\widetilde{t}_{1}$&$560$&$\widetilde{u}_{R}$&$1,055$&$m_{h}$&$121.4$\\ \hline
    $\widetilde{\chi}_{2}^{0}$&$217$&$\widetilde{\chi}_{2}^{\pm}$&$900$&$\widetilde{e}_{L}$&$571$&$\widetilde{t}_{2}$&$984$&$\widetilde{u}_{L}$&$1,146$&$m_{A,H}$&$973$\\ \hline
    $\widetilde{\chi}_{3}^{0}$&$896$&$\widetilde{\nu}_{e/\mu}$&$566$&$\widetilde{\tau}_{1}$&$109$&$\widetilde{b}_{1}$&$936$&$\widetilde{d}_{R}$&$1,096$&$m_{H^{\pm}}$&$976$\\ \hline
    $\widetilde{\chi}_{4}^{0}$&$899$&$\widetilde{\nu}_{\tau}$&$552$&$\widetilde{\tau}_{2}$&$561$&$\widetilde{b}_{2}$&$1,047$&$\widetilde{d}_{L}$&$1,148$&$\widetilde{g}$&$707$\\ \hline
                \end{tabular}
                \label{tab:masses}
\end{table}

There are two varieties of related selection cuts which we will entertain.  The original {\tt ULTRA} selections, so named for their emphasis of the ultra-high
jet multiplicity signal, are carefully documented in Ref.~\cite{Maxin:2011hy}.  These cuts are characterized by a rather low transverse momentum
threshold of $p_{\rm T} \ge 20$~GeV per jet, and require at least 150~GeV of missing energy $H_{T}^{\rm miss}$.  Any cut on the CMS collaboration
statistic $\alpha_{\rm T}$, which is designed to discriminate against false missing energy signals, is expressly forgone, as we have found
it to be likewise systematically biased against the ultra-high jet count signal.  Again, the high jet count threshold is in itself a quite 
effective discriminant against the SM background.  However, the {\tt ULTRA} selection cuts have a rather substantial drawback; specifically,
we have no deeply reliable estimates of the precise amount of background signal which does penetrate.  Although we have a baseline established
by substantial Monte Carlo of the crucial $t \bar{t} + {\rm jets}$ SM component, the massive cross-section of the QCD contribution makes a complete
proprietary simulation prohibitively difficult.

We thus also consider a second variety of selection cuts, which we refer to as the {\tt CMS HYBRID},
that has the distinct advantage of pre-tabulated SM background estimates from the CMS collaboration itself.  We describe the details
of this selection methodology in Refs.~\cite{Li:2011fu,Li:2011av}, where it is compared to the CMS report~\cite{PAS-SUS-11-003}.  By carefully extracting
just those events above the nine jet threshold from the detailed histogram plot presented in that reference, we estimate a SM background
of approximately $3.1 / {\rm fb}^{-1}$ events for the {\tt CMS HYBRID} selection cuts.  In most critical regards, and essentially in suppression
of the $\alpha_{\rm T}$ selection, these cuts are quite similar to our original {\tt ULTRA} preference, which is the reason for the ``hybrid''
description.  There are various incidental historical distinctions, including a reduction of the missing energy threshold to
$H_{T}^{\rm miss} \ge 100$~GeV, and the introduction of cuts on the electromagnetic fraction and an upper bound on the ratio of
missing energy carried by hard to soft jets.  The most notable difference is an increase in the transverse momentum required per
hard jet, to $p_{\rm T} \ge 50$~GeV.  While the greater overall jet count produced by the {\tt ULTRA} cuts is quite valuable for
discovery at the lighter end of the $\cal{F}$-$SU(5)$ model space (well below $M_{1/2} \simeq 500$~GeV or $m_{\rm LSP} \simeq 100$~GeV),
we have found~\cite{Li:2011av} that the greater background suppression of the {\tt CMS HYBRID} cuts becomes valuable for the lower-producing
heavier spectra.  Since our presently favored~\cite{Li:2011av} region of the model is near the just mentioned transition, where the discoverability
produced by the two cuts appears to come roughly into parity, we consider the {\tt CMS HYBRID} selections to be a very acceptable alternative.
Because of the clear advantage in quantification of the background competition, we opt to employ these cuts in the first tier of our
study, dealing with model discoverability.  We shall find in the second tier of this study that the large event counts required to drive a
fitting of the LSP mass will instead naturally favor the {\tt ULTRA} selections.

\begin{figure*}[htp]
	\centering
	\includegraphics[width=0.81\textwidth]{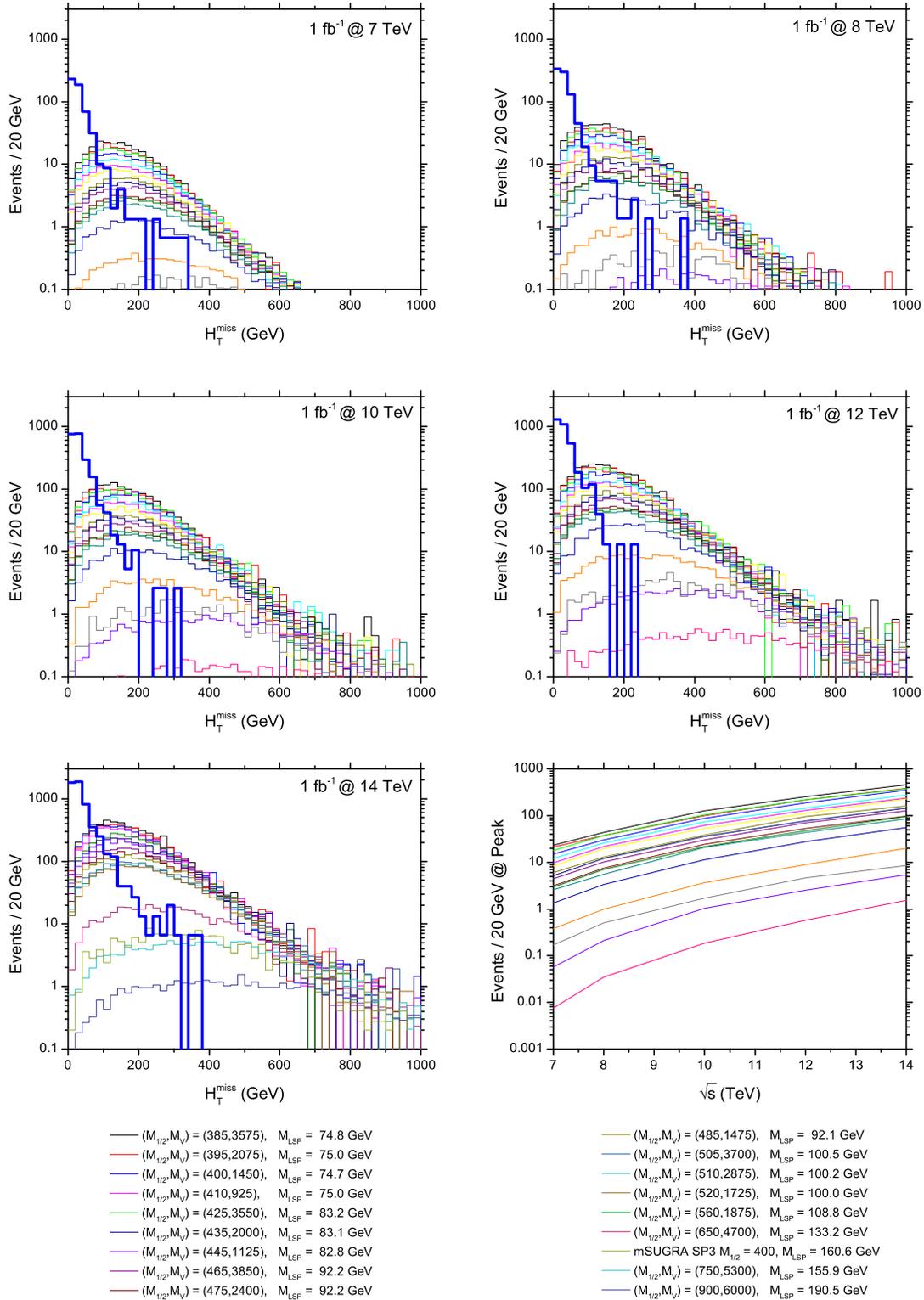}
	\caption{We depict the statistical distribution of missing energy events, per 20 GeV bin, scaled down to 1 ${\rm fb}^{-1}$ for LHC center-of-mass
	collision energies of $\sqrt{s} = (7,8,10,12~{\rm and}~14)$~TeV. Each frame displays the histogram count for 17 representative
	$\cal{F}$-$SU(5)$ benchmark points. The properties of the each individual point are listed in Table~\ref{tab:points}.
	Also included for comparison is the ``Snowmass Points and Slopes'' (SPS) benchmark point SPS3~\cite{Allanach:2002nj}. The thick
	blue histogram peaking at low energies represents the $t \overline{t} + {\rm jets}$ background sample. The lower right frame identifies the
	peak event count as a function of $H_{\rm T}^{\rm miss}$ for a continuous progression of LHC center-of-mass collision energies.
	The legend associates the gaugino mass $M_{1/2}$, vector mass $M_{V}$, and LSP mass $m_{LSP}$ of each of the 17 $\cal{F}$-$SU(5)$
	points and SPS3 point with its respective color in the diagrams.}
	\label{fig:MET_6plex}
\end{figure*}

The missing transverse energy which has been discussed is defined formally as follows.
\begin{equation}
H_{\rm T}^{\rm miss} \equiv
\sqrt{{\left( \sum_{\rm jets} p_{\rm T} \cos \phi \right)}^2 + {\left( \sum_{\rm jets} p_{\rm T} \sin \phi \right)}^2}
\label{HTmiss}
\end{equation}
Our simulations indicate a strongly peaked Poisson-like distribution of the histogram on missing transverse energy,
similar to the power spectrum of black body radiation with various temperatures. 
Figure Set~(\ref{fig:MET_6plex}) unmistakably highlights this physically analogous behavior by the way of five
histograms delineating the $\ge 9$ jets event rate as a function of $H_{\rm T}^{\rm miss}$, for LHC center-of-mass collision
energies of $\sqrt{s} = (7,8,10,12~{\rm and}~14)$~TeV, for each of the 17 representative $\cal{F}$-$SU(5)$ points
and the ``Snowmass Points and Slopes'' (SPS) benchmark point SPS3~\cite{Allanach:2002nj}. Curiously, the central value
of this distribution appears to correlate strongly with a small multiple of the LSP neutralino mass, a simple observation
which shall develop into the cornerstone of our second tier analysis strategy, to be presented subsequently.

All 2-body SUSY processes are included in Monte Carlo collider-detector simulation, which
has been performed using the {\tt MadGraph}~\cite{Stelzer:1994ta,MGME} suite, including
the standard {\tt MadEvent}~\cite{Alwall:2007st}, {\tt PYTHIA}~\cite{Sjostrand:2006za} and {\tt PGS4}~\cite{PGS4} chain. 
Post-processing has been performed by a script {\tt CutLHCO} of our own design (available for download~\cite{cutlhco}) that implements the desired cuts,
and counts and compiles the associated net statistics.  We likewise offer for download~\cite{lspanalysis} the pair of Mathematica
notebooks used to perform the primary LSP analysis and plot generation for the present paper.  The missing energy cut is relaxed for purposes
of plotting histograms in $H_{\rm T}^{\rm miss}$, and in particular for the second tier analysis to follow.  It is maintained, however, for the purposes
of our first tier signal discover analysis.  Our SUSY particle mass calculations have been performed using
{\tt MicrOMEGAs 2.1}~\cite{Belanger:2008sj}, employing a proprietary modification of the {\tt SuSpect 2.34}~\cite{Djouadi:2002ze}
codebase to run the RGEs.

%%%%%%%%%%%%%%%%%%%%%%%%%%%%%%%%%%%%%%%%%%%%%%%%%%%%%%%%%%%%%%%%%%%%%%%%%%%%

\subsection{The SUSY CDM Discovery Index}

%%%%%%%%%%%%%%%%%%%%%%%%%%%%%%%%%%%%%%%%%%%%%%%%%%%%%%%%%%%%%%%%%%%%%%%%%%%%

\begin{figure}[htp]
	\centering
	\includegraphics[width=0.45\textwidth]{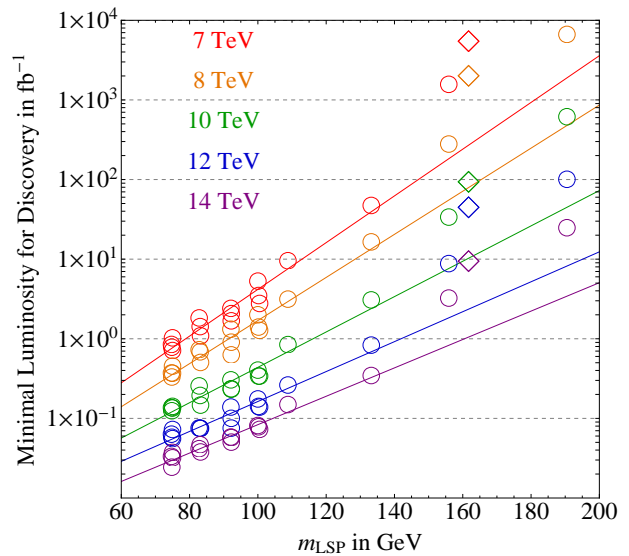}
	\caption{The projected minimal beam luminosity required to achieve a ratio $S/\sqrt{B+1} = 5$ of signal to background events
	with at least nine jets and at least $150$~GeV of missing transverse energy is plotted for collision
	energies of $\sqrt{s} = (7,8,10,12~{\rm and}~14)$~TeV, for each of the 17 representative
	$\cal{F}$-$SU(5)$ benchmark points.  Also included for comparison, with the diamond data markers, is the ``Snowmass Points and
	Slopes'' (SPS) benchmark point SPS3~\cite{Allanach:2002nj}.}
	\label{fig:fit_discovery_index}
\end{figure}

A well known indicator of the statistical significance of a collider signal in comparison to some competing background is
the ratio $S/\sqrt{B+1}$ of signal events $S$ to the square root of background events $B$, plus one.  The ``plus one''
limiter in the denominator is a sanity check in the case of extremely small backgrounds, which prevents the signal count
$S$ from being numerically smaller than the target ratio.  The conventional wisdom is that $S/\sqrt{B+1}\,\ge 5$ should
generally be considered favorable, with likewise a minimum of five signal events.
Scaling both $S$ and $B$ by the factor $N$, the quadratic equation may be readily inverted to establish the following relationship.
\begin{equation}
N = \frac{12.5\, B}{S^2} \times \left[ 1 + \sqrt{ 1 + {\left( \frac{2S}{5B} \right)}^2} \,\right]
\label{eq:discovery_index}
\end{equation}
In the large-background limit, where the ``plus one'' term is not required, this reduces to simply $25B/S^2$.
In Figure~(\ref{fig:fit_discovery_index}), we have graphically extrapolated the necessary scaling $N$ of integrated
luminosity which would be necessary to achieve the target ratio of $5$ for our discovery index.  Since we have
a well defined background count only for the current $\sqrt{s} = 7$~TeV beam, we have extrapolated the observed event
scaling for our own $\cal{F}$-$SU(5)$ Monte Carlo onto the backgrounds.  Averaging across the model space, and omitting the
heaviest two benchmarks (which scale somewhat more sharply than the primary spectra, demanding larger energies for any substantial
event production), we find event ratios of approximately $(1.0:2.3:8.3:21:41)$ for beam energies of $\sqrt{s} = (7,8,10,12,14)$~TeV.
As demonstrated in the plot, the logarithm of the scaling factor varies roughly linearly with the LSP mass, translating even
slight differences into dramatic distinctions in the prospects for discovery.  We likewise omit the heaviest two $\cal{F}$-$SU(5)$
benchmarks, as well as the mSUGRA control point, from the linearized fitting.  Incidentally, some portion of the visible deviation
from logarithmic linearity in the figure may be attributed to the ``plus one'' limiter.

To summarize the findings depicted in Figure~(\ref{fig:fit_discovery_index}), $1~{\rm fb}^{-1}$ of data at $\sqrt{s} = 7$~TeV could
be sufficient to unearth an LSP mass up to about $80$~GeV, while a $\sqrt{s} = 14$~TeV beam approximately doubles the discovery threshold
to around $160$~GeV, encompassing almost the entirety of No-Scale $\cal{F}$-$SU(5)$, for the same luminosity.  This is consistent
with our conclusions, published elsewhere~\cite{Li:2011fu,Li:2011av}, that the model space below $M_{1/2} \simeq 480$~GeV
and $m_{\rm LSP} \simeq 90$~GeV is disfavored by the recent CMS~\cite{PAS-SUS-11-003} and ATLAS~\cite{Aad:2011qa} reports on the
first femtobarn-plus of data, while the models near $M_{1/2} \simeq 520$ and $m_{\rm LSP} \simeq 100$~GeV are not only allowed, but well
described by the low-statistics excesses which have been observed.  Likewise, it is expected that the already collected $5~{\rm fb}^{-1}$ of data should be sufficient
for conclusive validation or refutation of that same model region.  For comparison, the discovery threshold for the SPS3 benchmark is overlaid
onto the same plot, with diamond shaped markers.  The prospects in this case are much less favorable than for the leading $\cal{F}$-$SU(5)$ benchmarks,
with a massive target luminosity of more than $5,000~{\rm fb}^{-1}$ at $\sqrt{s} = 7$~TeV suggested for discovery in this case.
The primary reason for this is simply that the LSP mass of the SPS3 example is quite heavy, at more than $160$~GeV.
This is not, however, incidental.  In fact, it is fairly generic that experimentally viable mSUGRA/CMSSM parameterizations
require rather heavy LSP particles in conjunction with a mass spectrum which is sufficiently globally heavy to evade the limits on
squark masses.  This in turn makes the discovery of a missing energy collider signal substantially less likely.
On top of this, the mSUGRA/CMSSM models generally underperform the No-Scale $\cal{F}$-$SU(5)$ models in the ultra-high
jet multiplicity regime, although it is difficult to make an absolute comparison, as the detailed spectra may be
hierarchically distinct, and not only different in overall scale.  In particular, we note that the example spectrum of
Table~\ref{tab:masses} should not be na\"ively compared against squark limits which are based on an assumed CMSSM/mSUGRA
construction; a preferable approach is the direct fitting of a Monte Carlo simulation against detailed collider measurements
which feature identically applied selection cuts~\cite{Li:2011fu,Li:2011av}.

It is unsurprising to note that detectability is improved dramatically as one looks beyond the current $\sqrt{s} = 7$~TeV LHC to
the future $\sqrt{s} = 14$~TeV.  Of course, the alternate path to achieving enhanced detectability, namely lightness of the
LSP mass, is one whose fundamental mechanism is outside of our hands.  It is, however, precisely the combination of a light LSP with
heavy squarks (excepting the light stop), again the central feature of the characteristic No-Scale $\cal{F}$-$SU(5)$ $m_{\tilde{t}_1} < m_{\tilde{g}} < m_{\tilde{q}}$
mass hierarchy, which lies at the heart of this particular model's best prospects for imminent discoverability.  The same model characteristic
is simultaneously responsible for the preponderance of ultra-high multiplicity jet events which may further facilitate that
discovery.  Given the relatively small target luminosities required by the intermediately heavy model space, there is moreover a
tangible possibility that the first delicate bouquet of a SUSY signal already perfumes the published data~\cite{Li:2011av},
or lacking appropriate selections, that it remains enveloped beneath the detritus of lower jet count SM events~\cite{Li:2011fu}.

At a time when the LHC, LEP, Tevatron and XENON100 results are strongly impinging upon the key parameter space of the CMSSM, and
many models which are favorable in terms LSP production are being ruled out by their squark spectra (which are likewise relatively light
across the board), the efficient simultaneous i) evasion of experimental limits and ii) preservation of
robust near term term testability represents no mean feat.  Although we have focused here
predominantly on collider-detector methodology, one should not meanwhile neglect the very
real possibility that the direct detection experiments such as XENON100~\cite{Aprile:2011hi} may trump the LHC to discovery
and/or classification of a SUSY CDM candidate~\cite{Li:2011in,Buchmueller:2011ki}.  We have remarked~\cite{Li:2011in},
in particular, that the dual designation of Supersymmetric Dark Matter in fact requires the contribution of both search
approaches, simultaneously complements and competitors.

%%%%%%%%%%%%%%%%%%%%%%%%%%%%%%%%%%%%%%%%%%%%%%%%%%%%%%%%%%%%%%%%%%%%%%%%%%%%

\section{Tier 2: SUSY CDM Classification}

%%%%%%%%%%%%%%%%%%%%%%%%%%%%%%%%%%%%%%%%%%%%%%%%%%%%%%%%%%%%%%%%%%%%%%%%%%%%

\subsection{The Boltzmann Analogy}

%%%%%%%%%%%%%%%%%%%%%%%%%%%%%%%%%%%%%%%%%%%%%%%%%%%%%%%%%%%%%%%%%%%%%%%%%%%%

The second tier of our analysis looks beyond discovery to model classification, and is focused on the detailed
spectroscopic distribution of the missing energy counts.
We were struck upon initial observation of the histogram counts against missing transverse energy
$H_{\rm T}^{\rm miss}$, as displayed in Figure Set~(\ref{fig:MET_6plex}), by the strong resemblance to the curves describing
the power spectrum of a black body radiator at various temperatures.  The memory stirs by this analogy
to recall the treasure of information enciphered in the cosmic microwave background which was
extracted by application of this single same key.  We have been most interested to further note the
rough visible correlation between the distribution peak ${(H_{\rm T}^{\rm miss})}^{\rm max}$,
and a small multiple of the LSP mass. Given the allowed kinetic excess over the jet invariant mass, the trigonometric reduction from
extraction of the beam-transverse component, the certainty of partial cancellation between the 
dual neutralino signal (with randomly oriented directionality due to the multi level decay cascade), and the possibility
of some substantial detector mismeasurement in at least one of the $\ge 9$ event jets, it does not at all seem obvious that
any such clear correlation should persist.

If one hopes nonetheless to reverse the correlation which is indeed observed, extrapolating from the statistics accrued under
actual collider-detector operational conditions to the active LSP mass, then it is essential to have
a quantitative understanding of the parameter dependencies in the event count function.  An empirical
reconstruction from the Monte-Carlo may serve as well as an analytically established relationship, if
one is careful to choose a physically motivated ans\"atz.  We will indeed here establish just such an empirical fitting relationship,
led by the Boltzmann analogy to consider a distribution of the Poisson type.  This single function will be demonstrated to simultaneously
describe the entirety of the $\cal{F}$-$SU(5)$ model space.  While the fitting function is not truly universal, insomuch as it fails to
adequately model the SPS3 control sample, we interpret this as an opportunity to distinguish experimentally between the
signatures of various candidate scenarios.  While the specific numerics may be quite model dependent, we speculate that the
underlying Poisson correlation between the dark matter particle mass and missing transverse energy may be fully general.

Most notably, we will successfully establish a specific numerical linkage between the LSP mass and the value of
${(H_{\rm T}^{\rm miss})}^{\rm max}$, with cumulative uncertainties totaling approximately $12-15\%$, which we
find to be quite satisfactory.  The co-linear relationship connecting the universal gaugino mass $M_{1/2}$ with the LSP
neutralino mass in No-Scale $\cal{F}$-$SU(5)$ must also be stressed, to the extent that precision measurement of the LSP mass
via a global fit of the $H_{\rm T}^{\rm miss}$ histogram is virtually equivalent to an experimental determination of $M_{1/2}$ itself.
As mentioned in Section~\ref{sct:MET}, the larger overall event multiplicities associated with the {\tt ULTRA} selection cut
procedure provide a much greater data density and uniformity than can be achieved with the {\tt CMS HYBRID} cuts, and are
thus preferable for use in our second tier study.  Again, the selection cut on $H_{\rm T}^{\rm miss}$ is relaxed for the
purposes of generating a distribution in this parameter.

%%%%%%%%%%%%%%%%%%%%%%%%%%%%%%%%%%%%%%%%%%%%%%%%%%%%%%%%%%%%%%%%%%%%%%%%%%%%

\subsection{An Empirical Poisson Fit\label{sct:lspfit}}

%%%%%%%%%%%%%%%%%%%%%%%%%%%%%%%%%%%%%%%%%%%%%%%%%%%%%%%%%%%%%%%%%%%%%%%%%%%%

A black body, in the Boltzmann limit, is described by a rescaling of the Poisson distribution
\begin{equation}
P = \frac{\mu^N e^{-\mu}}{N !} \, ,
\label{eq:poisson} 
\end{equation}
which represents an expansion of the binomial distribution in the limit of rare events, giving the
likelihood of making $N$ observations against an expected baseline of $\mu$ observations.  It is
important in this context, however, to note that the independent variable is taken as $\mu$ rather than
$N$, and the constant factorial denominator may then be replaced by a more general leading constant $A$
(not to be confused with the trilinear soft terms).
The differential black body intensity per frequency step is then described by
$(\mu \rightarrow \nu/T, N \rightarrow 3)$, or alternatively by
$(\mu \rightarrow 1/\lambda T, N \rightarrow 5)$ with respect to differential steps in the wavelength.

By trial, an excellent fit may be obtained for the frequency-style distribution, while the inverse wavelength-style distribution
seems much less well suited to model the simulation.  However, the non-zero event count in the limit $H_{\rm T}^{\rm miss} \rightarrow 0$ forces
one to consider (at least) a linear transformation of the horizontal axis, to $\mu = m \times (H_{\rm T}^{\rm miss}/ 100~{\rm GeV}) + b$, with
dimensionless slope and intercept parameters.  Such a relinearization of the Poisson distribution is familiar to
high-energy experimentalists, for example, as applied to the detection probability for a horizontal axis representing some
(potentially unknown) event cross section, with integrated beam luminosity as the slope parameter, and the background process
count representing the intercept.

The exponent $N=3$ is inessential to our purpose, and a sequential consideration of somewhat
larger integral values provided in several cases a comparable fitting quality;  nevertheless, with no single strongly favored
alternative, we will opt to respect the physical analogy by retention of the cubic power.  Together then, we adopt a three parameter
fitting function, written as follows, where the leading coefficient $A$ carries the unit $[1~{\rm fb} / 20~{\rm GeV}]$.
\begin{eqnarray}
\left( \frac{\rm Events}{20~{\rm GeV}\times 1~{\rm fb}^{-1}} \right) =
A \times \left[\,m \left( \frac{H_{\rm T}^{\rm miss}}{100~{\rm GeV}} \right) + b \,\right]^3 \times
\nonumber \\
\exp \left\{ - \left[\, m \left( \frac{H_{\rm T}^{\rm miss}}{100~{\rm GeV}} \right) + b \,\right]\right\} \quad
\label{eq:ansatz}
\end{eqnarray}
Explicitly differentiating the prior function, and setting the result to zero, one readily solves for the location of the
function maximum on the $H_{\rm T}^{\rm miss}$ axis.  Secondarily reinserting $(H_{\rm T}^{\rm miss})^{\rm max}$ back
into Eq.~(\ref{eq:ansatz}), an expression for the function maximum is likewise obtained.
\begin{eqnarray}
(H_{\rm T}^{\rm miss})^{\rm max} = 100~{\rm GeV} \times \left(\frac{3-b}{m}\right)
\label{eq:htmax} \\
{\left(\frac{\rm Events}{20~{\rm GeV}\times 1~{\rm fb}^{-1}}\right)}^{\rm max} =
A \times {\left(\frac{3}{e}\right)}^{3}
\label{eq:emax}
\end{eqnarray}
Integrating across  $H_{\rm T}^{\rm miss}$, one may establish the net event count per inverse femtobarn of luminosity,
a measure which corresponds to the overall model discoverability, as emphasized in the first tier of the present work.
The result is expressed via the partial Euler Gamma function, or may be polynomial expanded, as follows.
\begin{equation}
\left( \frac{\rm Events}{1~{\rm fb}^{-1}} \right) =
\left( \frac{100~{\rm GeV} \times 3!}{20} \right) \times \left( \frac{A\, e^{-b}}{m} \right) \times \sum_{i=0}^3 \frac{b^i}{i!}
\label{eq:net_events}
\end{equation}
A root-mean-square quantification $\sigma_{H_{\rm T}^{\rm miss}}$ of the Poisson function width may also be integrated in closed form.  The result will
simplify considerably once a numerical value is established for the intercept parameter $b$.
\begin{eqnarray}
\sigma_{H_{\rm T}^{\rm miss}}
\equiv \sqrt{ \langle {(H_{\rm T}^{\rm miss})}^2 \rangle - {\langle H_{\rm T}^{\rm miss} \rangle}^2}
= \left( \frac{100~{\rm GeV}}{m} \right) \times
\nonumber \\
\sqrt{\frac{b^6+12 b^5+66 b^4+192 b^3+288 b^2+288 b+144}{b^6+6 b^5+21 b^4+48 b^3+72 b^2+72 b+36}}\,\,\,
\label{eq:pwidth}
\end{eqnarray}

\begin{figure}[htf]
	\centering
	\includegraphics[width=0.45\textwidth]{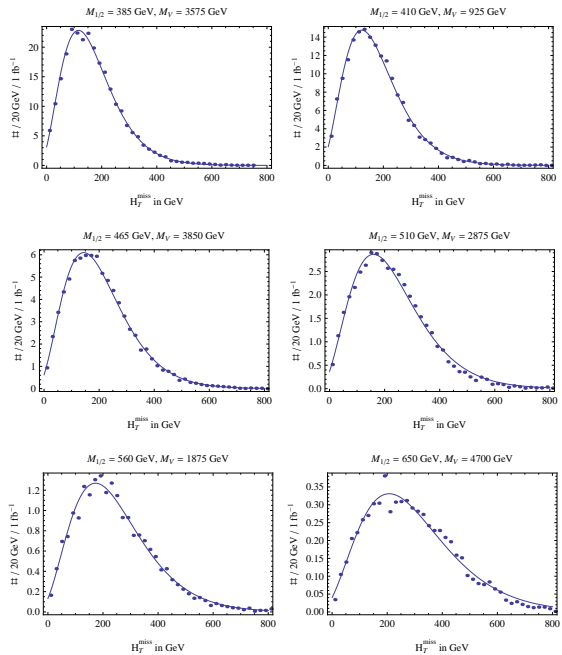}
	\caption{The counts of missing energy events, per 20 GeV bin, scaled down to 1 inverse femtobarn, are shown
	as blue dots from our Monte Carlo simulation.  Six representative benchmarks are selected for display,
	each at a collider beam energy of $\sqrt{s} = 7$~TeV.  The solid blue line is the best three parameter
	Poisson fit to each isolated model.}
	\label{fig:fit_poisson}
\end{figure}

Best fits, using a least-squares deviation metric, were composed for the parameters $(A,m,b)$, for each
of the complete Monte-Carlo event simulations of our seventeen $\cal{F}$-$SU(5)$ benchmarks, repeated
for each of the LHC center-of-mass collision energies $\sqrt{s} = (7,8,10,12,14)$~TeV.  Figure~(\ref{fig:fit_poisson}) depicts
a representative sample of six benchmarks, each at  $\sqrt{s} = 7$~TeV, demonstrating the broad suitability of this function.
We do find that beyond approximately $M_{1/2} = 650$~GeV, the low event cross-sections result in large detection fluctuations,
rendering a fit against the quantity of collected statistics somewhat tenuous.  We have therefore omitted the two benchmarks in
this category from the present analysis.

\begin{figure}[htf]
	\centering
	\includegraphics[width=0.45\textwidth]{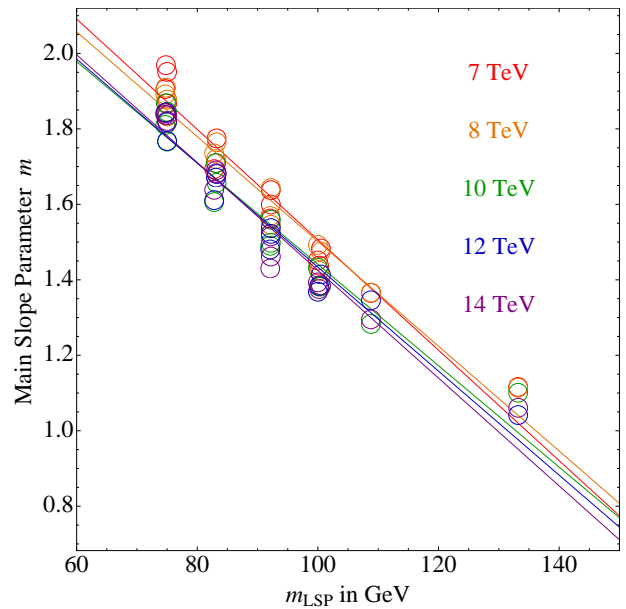}
	\caption{We depict the secondary linear fit of the slope parameter $m$ of each benchmark $\cal{F}$-$SU(5)$ model
	as a function of the LSP mass for each considered collider beam energy.}
	\label{fig:fit_main_slope}
\end{figure}

Given data for the three parameters of Eq.~(\ref{eq:ansatz}), across our full model space and for various collider energies,
the next task in the development of a suitably broad empirical function is the secondary functional correlation of each of
these parameters with the model and collider inputs.  We turn first to the main slope parameter $m$.  Figure~(\ref{fig:fit_main_slope})
demonstrates the strong linear correlation of $m$ with the LSP mass for fixed beam energy $\sqrt{s}$.  In fact, the basic beam energy
independence of the parameter $m$ is exhibited by the concurrency of the five overlaid color coded sample sets. 

An equally satisfactory fit for $m$ is possible against $M_{1/2}$, which is unsurprising as one expects the basic proportionality
$m_{\rm LSP} \propto M_{1/2}$, but since our purpose here is evocation of the LSP mass, we prefer the initially stated parameterization.
A weak secondary, though also apparently systematic, dependence of $m$ is observed against the vector-like mass $M_{\rm V}$ and the
top quark mass $m_{\rm t}$.  However, the statistical scatter is large, and we judge the additional complication to be unjustified
by the meager improvement in overall fit quality.  Given each of the prior facts, we settle on a final numerical fitting for the main
slope parameter $m$ which is given by averaging across the beam energy $\sqrt{s}$ for the five least-squares linearizations depicted in
Figure~(\ref{fig:fit_main_slope}).
\begin{equation}
m = -1.40 \times \left(\frac{m_{\rm LSP}}{100~{\rm GeV}}\right) + 2.86
\label{eq:fitm}
\end{equation} 
The quality of this fit may be characterized in a standard manner by the proportional deviation measure $\overline{\sigma}$,
which we notate with an overline to distinguish the relative scaling procedure, and define as shown following.
\begin{equation}
{\overline{\sigma}}_x \equiv \sqrt{\frac{\sum_{i=1}^{N}{\left( 1 - \frac{x_i^{\rm fit}}{x_i} \right)}^2}{N-1}}
\label{eq:propdev} 
\end{equation}
We find a proportional deviation in this case of ${\overline{\sigma}}_m = 0.043$, or crudely speaking, about $4 \%$ disagreement between the
single model fits for the slope parameter $m$ and the global numerical fitting for all models. 

\begin{figure}[htf]
	\centering
	\includegraphics[width=0.45\textwidth]{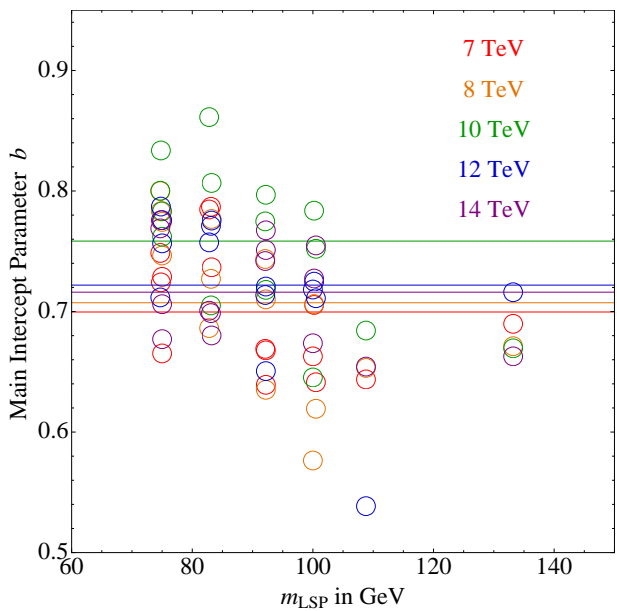}
	\caption{We depict the secondary averaging of the intercept parameter $b$ of each benchmark $\cal{F}$-$SU(5)$ model
	as a constant function of the LSP mass for each considered collider beam energy.}
	\label{fig:fit_main_intercept}
\end{figure}

Moving on to the main intercept parameter $b$, Figure~(\ref{fig:fit_main_intercept}) demonstrates substantially reduced linearity against the
LSP mass.  The situation is not improved for alternate potential independent fitting variables. Although there is a
recognizable systematic trend, there is a wholesale abandonment of the trend for certain sample points, and the
overall scatter remains large relative to the rise and fall of the best fit slopes.  We therefore opt in this case for a simple average,
as represented by the constant horizontal lines in the graphic.  Once again, there is no discernible beam energy dependence, and the average
of averages yields the following dimensionless best fit.
\begin{equation}
b = 0.721
\label{eq:fitb}
\end{equation}
Despite the overall appearance of statistical scatter, the clustering around the average is still sufficiently tight as a
relative measure to maintain a proportional deviation of only ${\overline{\sigma}}_b = 0.088$, representing disagreement on the level
of about $9 \%$ between the single model fits for the intercept parameter $b$ and the global all model average.

\begin{figure*}[htf]
	\centering
	\includegraphics[width=0.95\textwidth]{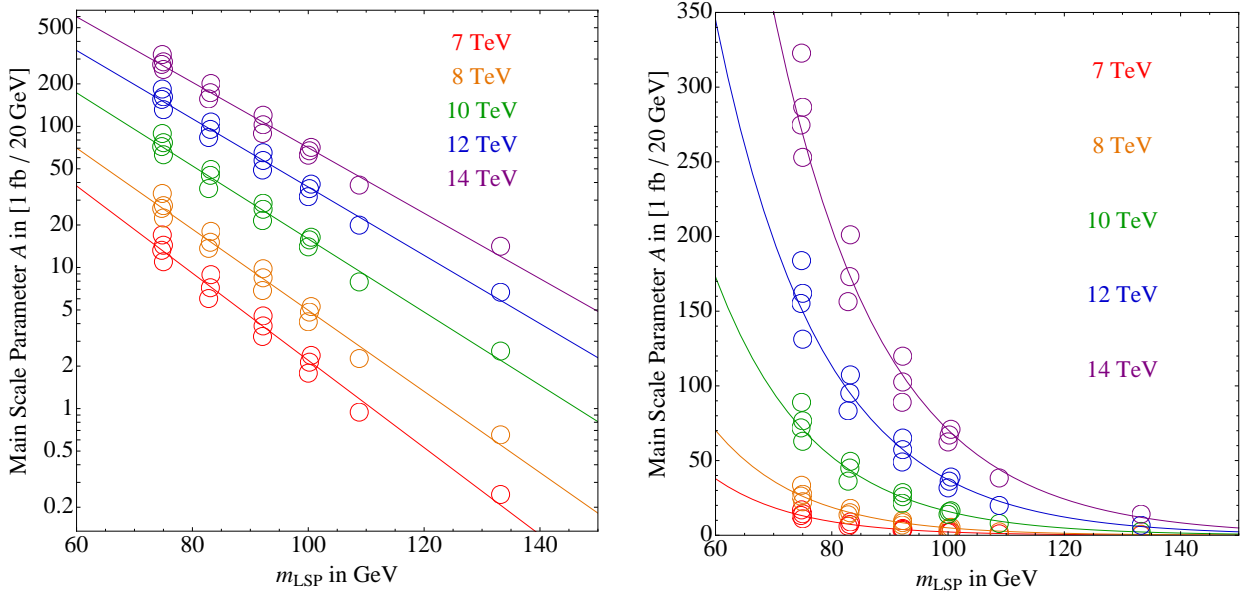}
	\caption{We depict the secondary exponential fit of the slope parameter $m$ of each benchmark $\cal{F}$-$SU(5)$ model
	as a function of the LSP mass for each considered collider beam energy.  The left-hand plot is logarithmic, while the right-hand
	plot is linear.}
	\label{fig:fit_main_scale}
\end{figure*}

Finally, we take up the secondary fit of the main scale parameter $A$ of Eq.~(\ref{eq:ansatz}), which is critical to
the overall model discoverability.  We observe, as attested to by
the first plot of Figure Set~(\ref{fig:fit_main_scale}) that it is the natural logarithm of $A \times ( 1~{\rm GeV} / 1~{\rm fb} )$
which scales linearly with the LSP mass,
rather than $A$ itself, implying an exponential dependence on the dimensionless parameter $(m_{\rm LSP}/100~{\rm GeV})$.  A fit of comparable
quality was obtained in a log-log fitting procedure, but the asymptotics of the power law scaling were deemed less physical.  The second
plot of Figure Set~(\ref{fig:fit_main_scale}), with the $A$ axis linearized, graphically demonstrates the incredibly large dynamic
range of the $A$ parameter across our model and LHC beam energy space, which spans a full three orders of magnitude.  It is apparent additionally 
in both plots that there is a tertiary functional dependence of the fit for $A$ against the LSP mass on the beam energy $\sqrt{s}$, particularly
in the fit intercept, but also in the slope.  This is to be expected, of course, insomuch as detection counts should scale rapidly with the
increasingly large interaction cross sections afforded by the increased phase space of a more energetic collider, as depicted in the lower
right frame of Figure Set~(\ref{fig:MET_6plex}), which identifies the peak event count as a function of $H_{\rm T}^{\rm miss}$
for a continuous progression of LHC center-of-mass collision energies.

\begin{figure*}[htf]
	\centering
	\includegraphics[width=0.95\textwidth]{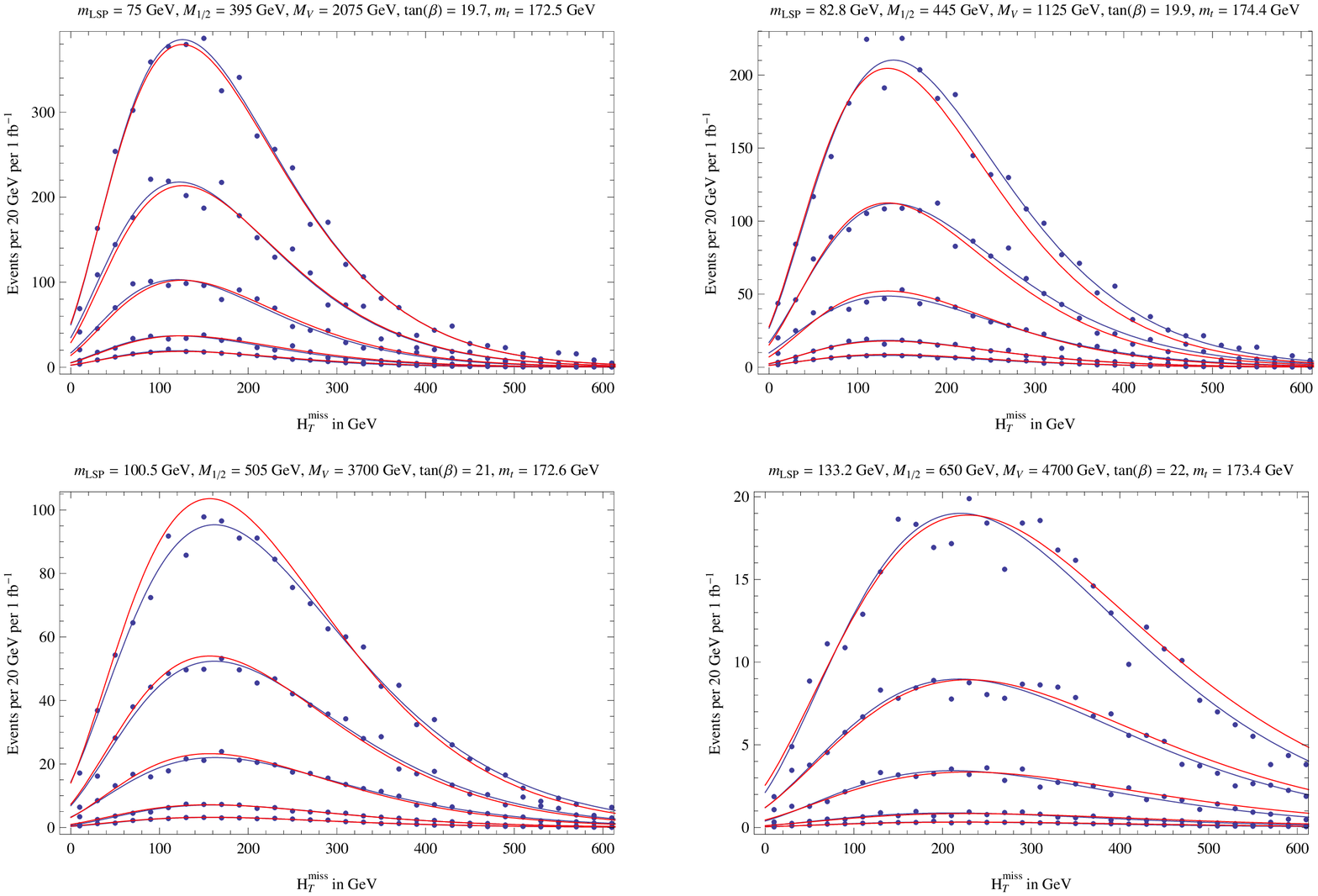}
	\caption{The counts of missing energy events, per 20 GeV bin, scaled down to 1 inverse femtobarn, are shown
	as blue dots from our Monte Carlo simulation.  Four representative $\cal{F}$-$SU(5)$ benchmarks are selected for display,
	for collider beam energies of $\sqrt{s} = (7,8,10,12~{\rm and}~14)$~TeV, from bottom to top.  The solid blue
	lines are the best three parameter Poisson fits to each isolated model.  The solid red lines are the global
	fit based only on the input parameters $(m_{\rm LSP},\sqrt{s},M_{\rm V},\tan \beta~{\rm and}~m_{\rm t})$.}
	\label{fig:fit_energies}
\end{figure*}

\begin{figure}[htf]
	\centering
	\includegraphics[width=0.5\textwidth]{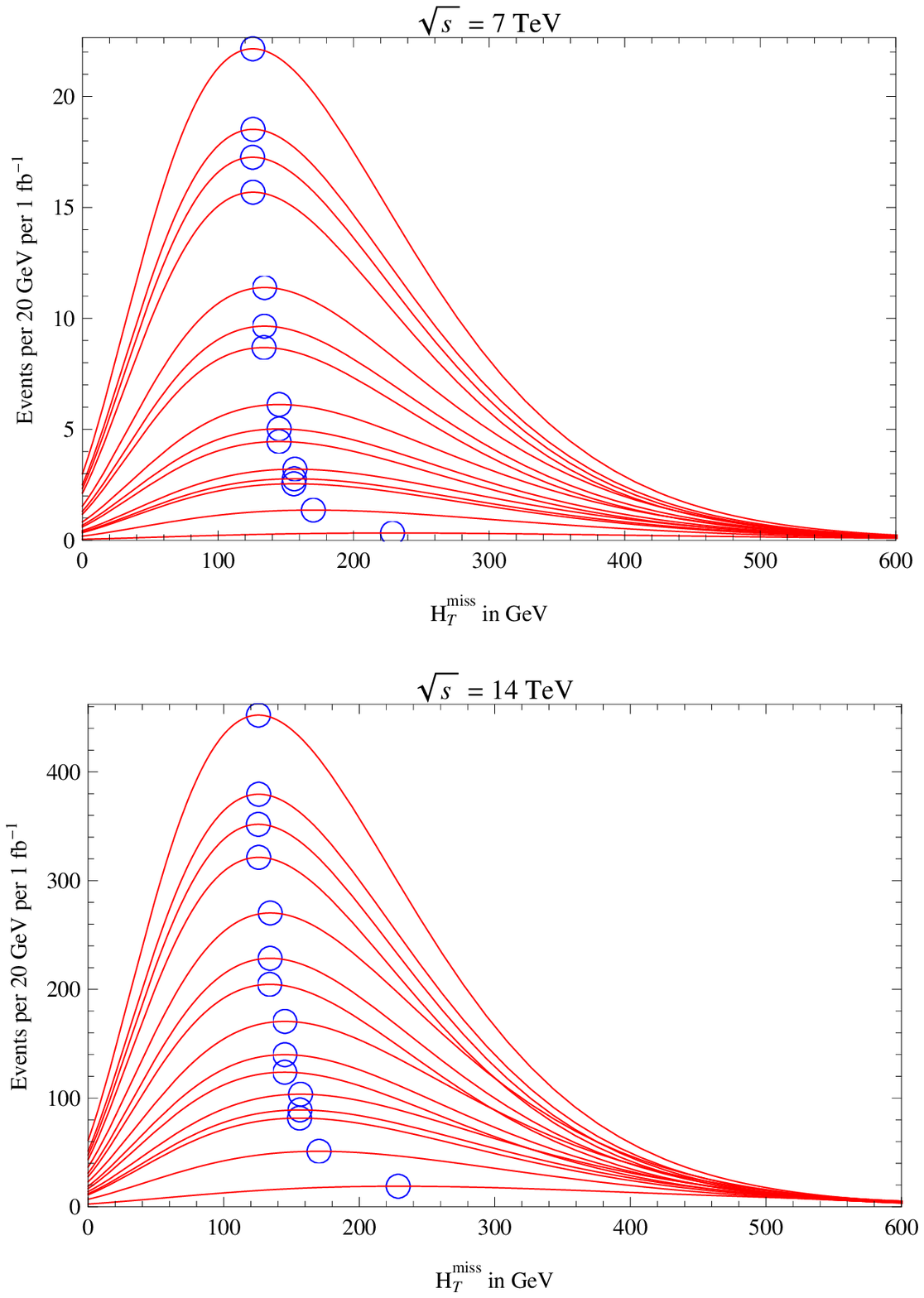}
	\caption{The global fit to missing energy is shown as a solid red line for each $\cal{F}$-$SU(5)$ benchmark model,
	with heavier spectra at the bottom, and lighter toward the top.  The blue rings denote the distribution peak.
	The plots are based only on the input parameters $(m_{\rm LSP},\sqrt{s},M_{\rm V},\tan \beta~{\rm and}~m_{\rm t})$,
	and not directly on the Monte Carlo. Collider beam energies of $\sqrt{s} = 7$~TeV and $\sqrt{s} = 14$~TeV are shown
	for the sake of comparison.}
	\label{fig:fit_benchmarks}
\end{figure}

The plots of Figure Set~(\ref{fig:fit_main_scale}) demonstrate moreover a substantial spreading, within a single beam energy contour,
for various benchmarks of similar LSP mass.  This fact, together with the extreme exponential scaling, suggest that if a suitable fitting
is possible, it must also take into consideration other input parameters such as $M_{\rm V}$, $\tan \beta$, and $m_{\rm t}$.
We omit $M_{1/2}$ from this list because of its strong co-linearity with $m_{\rm LSP}$.  Since the distribution peak and width are
unaffected by the overall scale term $A$, we are simultaneously less reluctant in this case to admit a somewhat more intricate
parameterization.  Guided by several additional plotting test cases for subordinate dependencies (the printing of which we here suppress),
a persistent eye for maximal functional simplicity and physicality, plus no small measure of trial and error, we settle at last on the
following form for the numerical value of $A$, in $[1~{\rm fb} / 20~{\rm GeV}]$ units.
\begin{eqnarray}
A&& = \exp \left[\, 20.0
-3.11 \times \left( \frac{m_{\rm LSP}}{100~{\rm GeV}} \right)
\right.
\nonumber \\
&& \phantom{} -2.51 \times \frac{\left( m_{\rm LSP} / 100~{\rm GeV} \right)}{\left( \sqrt{s} / 10~{\rm TeV} \right)}
-1.50 \times \left( \frac{10~{\rm TeV}}{\sqrt{s}} \right)
\nonumber \\
&& \phantom{} +0.861 \times \left( \frac{\sqrt{s}}{10~{\rm TeV}} \right)
+0.123 \times \left( \frac{M_{\rm V}}{1~{\rm TeV}} \right)
\nonumber \\
&& \left. \phantom{} -3.77 \times \left( \frac{\tan \beta}{20} \right)
-7.43 \times \left( \frac{m_{\rm t}}{174~{\rm GeV}} \right)
\,\right]
\label{eq:fitA}
\end{eqnarray}
The global fit against the main scale parameter $A$ is measured to have an extraordinarily small proportional deviation of
${\overline{\sigma}}_A = 0.036$, {\it i.e.}~less than $4 \%$.  With this final piece in place, the quality of the overall
fitting formula in $(m_{\rm LSP},\sqrt{s},M_{\rm V},\tan \beta~{\rm and}~m_{\rm t})$ which is expressed by the union of
Eqs.~(\ref{eq:ansatz},~\ref{eq:fitm},~\ref{eq:fitb} and \ref{eq:fitA}) may be judged by visual inspection of Figure
Set~(\ref{fig:fit_energies}).  Beam energies of $\sqrt{s} = (7,8,10,12~{\rm and}~14)$~TeV are there superimposed for each of
four representative benchmark samples.  The blue dots represent the histogram binning from the Monte Carlo, the blue curves
represent the least-squares fit to the single sample space in isolation, and the red curves represent the global fitting
function.  Figure Set~(\ref{fig:fit_benchmarks}) next depicts a scan across all benchmarks (excepting $M_{1/2} > 650$~GeV) for center
of mass beam energies of $\sqrt{s} = 7$~TeV and $\sqrt{s} = 14$~TeV.  The histogram data and individual sample fits are omitted
for visual clarity.  The blue rings denote the distribution peak.

%%%%%%%%%%%%%%%%%%%%%%%%%%%%%%%%%%%%%%%%%%%%%%%%%%%%%%%%%%%%%%%%%%%%%%%%%%%%

\subsection{The Once and Future LHC}

%%%%%%%%%%%%%%%%%%%%%%%%%%%%%%%%%%%%%%%%%%%%%%%%%%%%%%%%%%%%%%%%%%%%%%%%%%%%

Having established the global Poisson fitting function, we now pause to reflect on the key purpose of this second
tier study, granting for argument's sake an imminent SUSY collider signal observation, and looking by the mind's eye
a bit farther ahead, perhaps beyond the present LHC to a forthcoming $\sqrt{s} = 14$~TeV incarnation, and a time after the
initial excitement of the SUSY CDM discovery has given way to the business of extracting the distinguishing details of the physical SUSY projection.

We consider first the parameter subset of just $m$ and $b$, as they are alone sufficient to fully characterize the location of the Poisson peak
in $H_{\rm T}^{\rm miss}$ from Eq.~(\ref{eq:htmax}), as well as the distribution width of Eq.~(\ref{eq:pwidth}). Substituting back into each of these relations,
we obtain the following, which are expected to apply across the full $\cal{F}$-$SU(5)$ model space.
\begin{eqnarray}
(H_{\rm T}^{\rm miss})^{\rm max} &\Rightarrow& \frac{228~{\rm GeV}}{m} \Rightarrow \frac{163~{\rm GeV}}{(2.04-\frac{m_{\rm LSP}}{100~{\rm GeV}}) } \quad
\label{eq:htmax2} \\
\sigma_{H_{\rm T}^{\rm miss}} &\Rightarrow& \frac{199~{\rm GeV}}{m} \Rightarrow \frac{142~{\rm GeV}}{(2.04-\frac{m_{\rm LSP}}{100~{\rm GeV}}) } \quad
\label{eq:pwidth2}
\end{eqnarray}
The expression from Eq.~(\ref{eq:net_events}) for the net event count per inverse femtobarn of luminosity similarly reduces as shown following,
where  $(m,A)$ carry the functional dependencies outlined in Eqs.~(\ref{eq:fitm},\ref{eq:fitA}).
\begin{equation}
\left( \frac{\rm Events}{1~{\rm fb}^{-1}} \right) \Rightarrow 29.8 \times \left( \frac{A}{m} \right) 
\label{eq:net_events2}
\end{equation}
The overall proportional deviation from the actual Monte Carlo counts is less than $1\%$, using the local $(A,m,b)$ parameters,
and still less than $5\%$ using the global fitting parameters.

Inverting Eq.~(\ref{eq:htmax2}), we can get the desired expression for the LSP mass in terms of the observed location of the
Poisson peak, or alternatively along with Eq.~(\ref{eq:htmax}), in terms of the Poisson fitting parameters.
\begin{eqnarray}
m_{\rm LSP}
&\Rightarrow& 204~{\rm GeV} - 163~{\rm GeV}\times\left(\frac{100~{\rm GeV}}{(H_{\rm T}^{\rm miss})^{\rm max}}\right)
\nonumber \\
&=& 204~{\rm GeV} - 163~{\rm GeV}\times \left(\frac{m}{3-b}\right)
\label{eq:mlsp}
\end{eqnarray}
Comparing our input LSP masses, as processed through the collider-detector Monte Carlo simulation, to the value obtained
by fitting the resulting $H_{\rm T}^{\rm miss}$ distribution according to the Eq.~(\ref{eq:ansatz}) ans\"atz, and passing the fit
parameters through Eq.~(\ref{eq:mlsp}), we observe a proportional deviation of ${\overline{\sigma}}_{m_{\rm LSP}} = 0.058$, or
about $6 \%$ error, induced by the closed circuit.  In a realistic application, this would of course necessarily be compounded
with any uncertainty in the fitting procedure itself.

Uncertainty in the overall fit against data comes from multiple sources.  First, there are statistical fluctuations and detector errors
in the measurements of the $H_{\rm T}^{\rm miss}$ histogram itself.  These should be reasonably modeled in our approach, which again
includes a post processing phase by {\tt PGS4}~\cite{PGS4} to simulate the necessarily faulty and incomplete nature of real world data.
However, it should be noted that although our histograms are normalized to $1~{\rm fb}^{-1}$ of data, we have generally overrun the simulation by some
(possibly large) factor and scaled down, which will tend to reduce random scatter.  Secondly, it is not {\it a priori} obvious that Eq.~(\ref{eq:ansatz})
itself carries any fundamental validity, although the fitting exemplified in Figure~(\ref{fig:fit_poisson}) does speak strongly for its use.  In an attempt to
quantify these two potential sources of error, we computed, for each $\cal{F}$-$SU(5)$ model and beam energy combination, the
absolute standard deviation between the Monte Carlo bin heights and the best fit Poisson curve of the given model, dividing at the
end by the average bin height, and finally averaging across the full model space.  An error of about $12 \%$ was observed.  Using
instead the global fit of Eqs.~(\ref{eq:fitm},\ref{eq:fitb}) to generate the $(m,b)$ parameters based on the known LSP mass, the
error increased to about $15 \%$.

An additional, and potentially more problematic, source of real world error, which we have easily sidestepped in simulations, is
the difficulty of deconvolving the SM background distribution from the SUSY signal.
We argue in Refs.~\cite{Maxin:2011hy,Li:2011hr} that the substantially elevated jet cut of greater than
or equal to nine jets overwhelmingly compromises the observability of the SM signal.  However, the footprint of the new physics is itself
sufficiently narrow that our simulation of the the $t \overline{t} + {\rm jets}$ background (in addition to the QCD and vector boson plus jets components
which we have not personally modeled, but which are accounted by the large collaborations such as CMS)
can generate a sufficient number of events to encroach upon the $\cal{F}$-$SU(5)$ signal.
As exhibited by the $t \overline{t}$ distribution in Figure Set~(\ref{fig:MET_6plex}), the SM background contributions may be expected to peak at lower energies,
and a hard lower bound cut on $H_{\rm T}^{\rm miss}$ is thus quite effective if one is interested in the integrated signal, for the pure initial
purposes of a proof of signal existence.  In the present case, however, it is quite troublesome that the missing energy $(H_{\rm T}^{\rm miss})^{\rm max}$
peak should occur in the vicinity of $100$~GeV, potentially shrouded and obscured by the SM veil.  

However, there is some substantial comfort available from the explicit analytic functional form of Eq.~(\ref{eq:ansatz}).  If one
tacitly accepts this ans\"atz distribution, then the behaviors at the peak and distance from the peak are implicitly coupled.  In fact,
it is possible to fully extrapolate the precise form of the three parameter fit in the region where background competes strongly
only by carefully observing the distribution in the region with low background.  Such an approach, combined certainly with a careful
subtraction of the estimated SM background at smaller $H_{\rm T}^{\rm miss}$ values, may present a surprisingly direct and well resolved
window on the mass of the underlying LSP particle which is responsible for the missing energy signal.  The essential decoupling of the simple
Eq.~(\ref{eq:mlsp}) result from the collider beam energy seems to us quite remarkable.  It is also worth remarking that the Eq.~(\ref{eq:fitA})
expression of the scale parameter $A$, while not so formally simple as the expression for the slope parameter $m$, does contain corroborating
information on the LSP mass, which is in some sense orthogonal to that extracted in Eq.~(\ref{eq:mlsp}).

%%%%%%%%%%%%%%%%%%%%%%%%%%%%%%%%%%%%%%%%%%%%%%%%%%%%%%%%%%%%%%%%%%%%%%%%%%%%

\subsection{The General and The Proprietary}

%%%%%%%%%%%%%%%%%%%%%%%%%%%%%%%%%%%%%%%%%%%%%%%%%%%%%%%%%%%%%%%%%%%%%%%%%%%%

\begin{figure}[htf]
	\centering
	\includegraphics[width=0.5\textwidth]{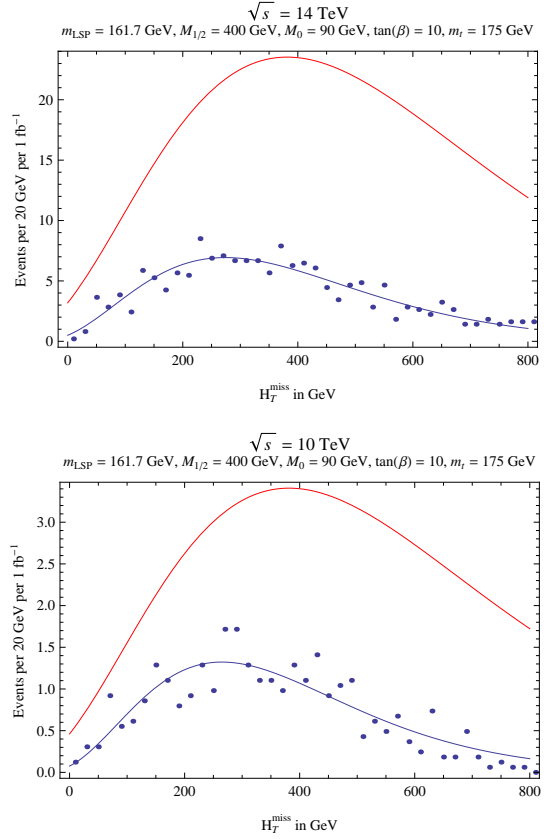}
	\caption{Statistical distributions of missing energy events, per 20 GeV bin, scaled down to 1 ${\rm fb}^{-1}$ for LHC center-of-mass
	collision energies of $\sqrt{s} = (7~{\rm and}~14)$~TeV, for the mSUGRA benchmark SPS3. Each frame displays the histogram count
	as blue dots from our Monte Carlo simulation.  The solid blue line is the best three parameter Poisson fit for this isolated model.
	The global $\cal{F}$-$SU(5)$ fit to missing energy is shown as a solid red line for the corresponding parameters
	of the SPS3 benchmark, with $M_{\rm V} = 0$.}
	\label{fig:fit_cmssm}
\end{figure}

In Figure Set~(\ref{fig:fit_cmssm}), we present the parallel analysis of our SPS3 mSUGRA control sample.  The actual histogrammed events (blue dots),
along with the blue line best fit, are contrasted with the corresponding global $\cal{F}$-$SU(5)$ fit, using the SPS3 parameters, with $M_{\rm V} = 0$.
The divergence is clear, in both scale and location of the distribution peak, immediately demonstrating that the empirical fit does not
directly generalize into alternative model spaces.
Nevertheless, the essential Poisson characteristic appears to be persistent, and we speculate that this bulk feature may be broadly applicable,
while the specific numerical coefficients may be model dependent.  There is also a second sense in which the ``tier 2'' procedure
may be general in principle, though again proprietary in practice.  This is with respect to the prospects for overcoming competing backgrounds.
The ultra-high jet multiplicity cutting procedure, which we suggest is capable of sufficiently elevating the No-Scale $\cal{F}$-$SU(5)$
$H_{\rm T}^{\rm miss}$ distribution above the SM background that a function fit might be
possible, may not be as effective for generic mSUGRA/CMSSM samples, which may moreover suffer from a necessarily heavier LSP candidate. 
We must also here reemphasize two key points: (1) Measuring the LSP neutralino mass is essentially equivalent to measuring
the universal gaugino mass $M_{1/2}$ in No-Scale $\cal{F}$-$SU(5)$ because the overall mass rescaling of the model is based only
on this single dimensionful GUT scale parameter; (2) If our method for the LSP neutralino mass determination
cannot be broadly applied outside of No-Scale $\cal{F}$-$SU(5)$, the absence of this rescaling property may be the deep reason.

One should be cautious in attempting to extend the discrepancy manifest by the global fitting function in Figure Set~(\ref{fig:fit_cmssm}).  For
example, the overage of the red line fit, which might casually be interpreted to imply a generically weaker discovery potential for the mSUGRA
candidates, can easily be converted to a severe underestimate by the transformation of $\tan \beta$ from 10 to 30, a switch which roughly
corresponds to adoption of the SPS1b Snowmass benchmark, rather than SPS3.  In this example, the strong dependence of Eq.~(\ref{eq:fitA}) on the ratio
$\tan \beta$, which is softened in $\cal{F}$-$SU(5)$ by a generic proximity to the numerical value of $20$, is exaggerated for the SPS models.
The lesson here is that comparisons of this sort are subtle, and that the conventional wisdom
developed in one model space does not necessarily transfer intact to an alternate class of constructions.  We remark, however, that the clear 
model dependency of the applicable Poisson fitting function, itself contains essential information, which may ultimately prove helpful
in the process of distinguishing between contending efforts to explain the ongoing accumulation of LHC data.

As a further control on the cross-applicability of our results, we fully regenerated the primary LSP fitting analysis presented in Section~\ref{sct:lspfit}
for two related but distinct scenarios, i) substituting the {\tt CMS HYBRID} cutting methodology, and ii) retaining the {\tt ULTRA} selections, but
substituting a $k_{t}$ jet clustering algorithm (specifying an angular distance parameter of $0.4$) for our default cone algorithm (these being the two
analysis options available within the {\tt PGS4}~\cite{PGS4} detector simulation).  While the event count scale, as characterized by our parameter $A$,
is affected in both cases, we find the correlation between $m_{\rm LSP}$ and the maximum of the the event histogram in $H_{\rm T}^{\rm miss}$, as
characterized by our parameters $m$ and $b$, to be rather stable.  Moreover, the leading modification to the scale parameter $A$ is a constant
multiplicative factor.  This is a rather reassuring result, which preserves the correlation established by Eq.~(\ref{eq:mlsp}),
which is again likewise independent of the collider beam energy.  We find it quite satisfactory that the overall event scale, which is the most
difficult parameter to fit, and the most sensitive to incidental variation of the experimental technique, is simultaneously the least critical
carrier of actual information on the $\cal{F}$-$SU(5)$ LSP particle and SUSY spectrum.

%%%%%%%%%%%%%%%%%%%%%%%%%%%%%%%%%%%%%%%%%%%%%%%%%%%%%%%%%%%%%%%%%%%%%%%%%%%

\section{Conclusions}

%%%%%%%%%%%%%%%%%%%%%%%%%%%%%%%%%%%%%%%%%%%%%%%%%%%%%%%%%%%%%%%%%%%%%%%%%%%%

The LHC era is rapidly evolving, with each major experimental collaboration having already accumulated a reported $5~{\rm fb}^{-1}$ of integrated
luminosity at the close of 2011.  Looking into the not so distant future, we must plan to focus not only on initial attainment of
the much anticipated SUSY signal discovery, but also on mining the wealth of information present in the data for clues to 
the reconstruction of a detailed SUSY mass spectrum. We have concentrated in this presentation on precisely such a two tiered objective. Firstly, we
have comprehensively documented the prospects for discovery of the full No-Scale $\cal{F}$-$SU(5)$ model space at various LHC center-of-mass energies,
based on the raw count of events with greater than $100$~GeV of missing transverse energy and at least nine jets.
Secondly, upon observing the dual curiosities of i) a material correlation between the $H_{\rm T}^{\rm miss}$ histogram peak and a small multiple of the LSP mass,
and ii) the strikingly familiar shape of the missing energy distribution, reminiscent of the power spectrum of a black body radiator at various
temperatures, we have implemented a broad empirical fit of our extensive Monte Carlo $H_{\rm T}^{\rm miss}$ simulation against a Poisson distribution
ans\"atz.  We advanced the resulting fit as a theoretical blueprint for deducing the mass of the LSP, utilizing only the missing transverse energy
in a statistical distribution of $\ge$ 9 jet events.  We judge the error directly associated with this technique to be within the satisfactory range of
$12-15\%$.  It will be interesting to see whether this supersymmetry mass extraction instrument might find some value by inclusion in
the experimentalist toolbag, as we journey into and beyond the presumed discovery phases of the LHC operation in the next two years.

We have presented this analysis in the context of the No-Scale ${\cal F}$-$SU(5)$ model, distinguished by its essential phenomenological consistency,
profound predictive capacity, singularly distinctive experimental signature, and imminent testability.  The core of the model space around and
above $M_{1/2} = 500$~GeV and $m_{\rm LSP} = 100$~GeV remains experimentally unblemished, in stark contrast to the incisive cuts already made into the
heart of the more traditional mSUGRA/CMSSM formulations.  This deft evasion of all existing LEP, Tevatron, and LHC sparticle mass limits stems from the
characteristic sparticle mass hierarchy $m_{\tilde{t}_1} < m_{\tilde{g}} < m_{\tilde{q}}$ of a light stop and gluino, both comfortably lighter than all other
squarks, a feature which is stable across the full model space.  With some reports claiming that less than one percent of the formerly viable mSUGRA and
CMSSM parameter space has withstood even the earliest accumulation of LHC data, the existence of a clearly defined, deeply theoretically motivated model
which simultaneously exhibits consistency with all current experiments alongside testability
at near term future experiments, is of some pressing interest.  The seventeen example spectra which we have studied span the model space of
bare minimal constraints on No-Scale $\cal{F}$-$SU(5)$, and establish a material correlation between the mass of the dark matter particle,
taken to be the LSP neutralino, and the collider missing energy signal.  It appears that the stability of the SUSY mass hierarchy, with the
overall scale driven by a single dimensionful parameter at the GUT scale, may be responsible for facilitating the remarkably robust
and simple fitting relationship which has been observed.  Together, these facts speak efficiently to the suitability of No-Scale $\cal{F}$-$SU(5)$
as a leading contender in the race to describe post Standard Model physics at the dawn of the LHC era. 

%%%%%%%%%%%%%%%%%%%%%%%%%%%%%%%%%%%%%%%%%%%%%%%%%%%%%%%%%%%%%%%%%%%%%%%%%%%%

\begin{acknowledgments}
This research was supported in part 
by the DOE grant DE-FG03-95-Er-40917 (TL and DVN),
by the Natural Science Foundation of China 
under grant numbers 10821504 and 11075194 (TL),
by the Mitchell-Heep Chair in High Energy Physics (JAM),
and by the Sam Houston State University
2011 Enhancement Research Grant program (JWW).
We also thank Sam Houston State University
for providing high performance computing resources.
\end{acknowledgments}

%%%%%%%%%%%%%%%%%%%%%%%%%%%%%%%%%%%%%%%%%%%%%%%%%%%%%%%%%%%%%%%%%%%%%%%%%%%%

\bibliography{bibliography}

\end{document}